\documentclass[twocolumn,amsfonts,aps,prd,nofootinbib,showpacs,10pt]{revtex4-1}
\usepackage{textcomp}
\usepackage{amsmath}
\usepackage{amsfonts}
\usepackage{calc}
\usepackage{mathrsfs}
\usepackage{amssymb}
\usepackage{amsmath}
\usepackage{array}
\usepackage{color}
\usepackage{bm}
\usepackage{graphicx}
\usepackage{hyperref}

\newcommand{\btilde}{\bm\tilde}
\renewcommand{\P}{\mathcal{P}}
\newcommand{\res}{\mathcal{R}}
\renewcommand{\S}{{\rm S}}
\newcommand{\R}{{\rm R}}
\newcommand{\e}{\epsilon}

\begin{document}
\title{Conservative effect of the second-order gravitational self-force on quasicircular orbits in Schwarzschild spacetime} 
\author{Adam Pound} 
\affiliation{Mathematical Sciences, University of Southampton, Southampton, United Kingdom, SO17 1BJ}
\date{\today}

\begin{abstract}
A compact object moving on a quasicircular orbit about a Schwarzschild black hole gradually spirals inward due to the dissipative action of its gravitational self-force. But in addition to driving the inspiral, the self-force has a conservative piece. Within a second-order self-force formalism, I derive a second-order generalization of Detweiler's redshift variable, which provides a gauge-invariant measure of conservative effects on quasicircular orbits. I sketch a frequency-domain numerical scheme for calculating this quantity. Once this scheme has been implemented, its results may be used to determine high-order terms in post-Newtonian theory and parameters in effective-one-body theory.
\end{abstract}
\maketitle 

\section{Introduction}
The gravitational self-force program was initiated with the goal of modeling extreme-mass-ratio inspirals (EMRIs)~\cite{Mino-Sasaki-Tanaka:97}, astrophysical systems in which stellar-mass compact objects spiral into far more massive black holes in galactic nuclei. An EMRI evolves primarily due to dissipation: the object emits gravitational waves that carry away energy (or equivalently, the self-force does negative work), causing the orbit to shrink until the object plunges into the black hole. However, in the years since the program began, the \emph{conservative} effects of the self-force have also proven to be a fecund area of study. These conservative effects must be accounted for to obtain accurate long-term models of inspirals~\cite{Pound-Poisson-Nickel:05,Pound-Poisson:08,Hinderer-Flanagan:08}, and their influence on long-term orbital evolution has recently been calculated concretely for the first time~\cite{Warburton-etal:12}. 

Besides its long-term effect on inspirals, the conservative piece of the self-force also tells us about short-term effects~\cite{Barack-Sago:07,Detweiler:08,Barack-Sago:09,Barack-Sago:10,Barack-Sago:11,Shah-etal:12}. The most obvious example might be a correction to the standard relativistic precession of an eccentric orbit. But conservative effects arise even in the case of quasicircular orbits (i.e., orbits that would be precisely circular in the absence of dissipation). For example, the radial force alters the frequency of an orbit at a given orbital radius.

Because quantities such as (coordinate) azimuthal angle and radius---and the gravitational self-force itself---are gauge dependent~\cite{Barack-Ori:01}, effects such as precession and frequency shifts at a given coordinate radius are as well. Hence, a primary goal when calculating self-force effects is to identify some gauge-invariant characterization of them. For example, orbital precession can be written in an invariant form in the circular limit~\cite{Barack-Sago:11}. A shift in frequency is invariant if the radius is physically identifiable; for example, one can consider the shift in frequency of the innermost stable circular orbit (ISCO)~\cite{Barack-Sago:09,Favata:11,LeTiec-etal:12b,Isoyama-etal:14}. For quasicircular orbits away from a special orbital radius, the principal invariant quantity of interest has been Detweiler's redshift variable, the inverse of the time component of a certain normalized four-velocity, which for later purposes I will denote by $\btilde u^t$~\cite{Detweiler:08}. The construction of this quantity is based on the fact that the orbit, which is accelerated by the self-force when considered to move in the background metric of the large black hole, is a geodesic when considered to move in a certain \emph{effective} metric, a certain smooth piece of the full, physical metric of the binary~\cite{Detweiler-Whiting:03,Pound:10a,Harte:12}. $\btilde u^t$ describes the ratio of proper time of an inertial observer at infinity to proper time along the orbit as measured in that effective metric. Its inverse, $1/\btilde u^t$, is the redshift experienced in the effective metric by a photon emitted to infinity in a direction perpendicular to the orbital plane. It can also be heuristically interpreted as the orbital energy as measured in a frame that co-rotates with the orbit. Because these interpretations of $\btilde u^t$ refer to quantities in the effective metric, rather than the binary's physical metric, their physical meaning is somewhat hazy. Nevertheless, defined strictly as the ratio of two measures of time, the quantity $\btilde u^t$ is invariant. Furthermore, it can be used to find other physical effects, such as the ISCO shift in Schwarzschild~\cite{LeTiec-etal:12b} and Kerr~\cite{Isoyama-etal:14}.

Invariant conservative quantities such as these are important beyond their role in characterizing the physics of extreme-mass-ratio binaries. They have been the point of comparison between self-force calculations performed in different gauges~\cite{Sago-Barack-Detweiler:08,Dolan:13}. More notably, in efforts originally led by Detweiler, Blanchet, and collaborators~\cite{Detweiler:08,Blanchet-etal:10a,Blanchet-etal:10b}, they have allowed for comparisons with entirely distinct models such as full numerical relativity and post-Newtonian (PN) theory~\cite{LeTiec-etal:11,Favata:11,LeTiec-etal:12b,LeTiec-etal:13,Shah-etal:13}. Since self-force calculations offer the only highly accurate model in the domain of extreme mass ratios and highly relativistic fields, they can also do better than compare: they set benchmarks for numerical relativity, and they have been used to determine high-order parameters~\cite{Blanchet-etal:10b,Favata:11,Shah-etal:13, Damour:09, Barack-Damour-Sago:10,Barausse-etal:11,Akcay-etal:12,Bini-Damour:13} in PN theory and the effective-one-body theory (EOB) introduced in Refs.~\cite{Buonanno-Damour:99,Damour-etal:00}. Furthermore, study of these conservative effects has provided strong evidence that the domain of validity of the self-force formalism can be made much larger than one would naively expect, pushing it toward modeling binaries of comparable-mass objects~\cite{LeTiec-etal:11,LeTiec-etal:12b,LeTiec-etal:13}.

Until recently, all of this work had been limited to linear order in the binary's mass ratio. Although some analyses had been performed at second order~\cite{Rosenthal:06a,Rosenthal:06b, Pound:10a, Detweiler:12}, they did not provide a practical means of concretely calculating second-order effects. However, with the recent development of complete second-order self-force formalisms~\cite{Pound:12a,Gralla:12,Pound:12b}, there is now no substantial obstacle to performing such concrete calculations. Proceeding to second order offers several exciting prospects: highly accurate calculations of effects on intermediate-mass-ratio and even comparable-mass binaries; stronger benchmarks for numerical relativity; and further improvements of the accuracy of PN and EOB models. The purpose of this paper is to take the first step toward realizing those goals. Restricting my attention to the simplest case, that of quasicircular orbits in Schwarzschild, I derive a gauge-invariant formula for a second-order generalization of Detweiler's redshift variable. I then outline how that quantity can be calculated numerically in the frequency domain.

\subsection{Plan of this paper}
Due to the nonlinear nature of the problem, defining and extracting conservative dynamics from a dissipating system at second order is more delicate than it was in the linearized problem. At first order, the time-symmetric part of the retarded solution was equal to the half-retarded-plus-half-advanced solution, and the force in the half-retarded-plus-half-advanced solution was equal to the conservative piece of the force in the retarded solution. At second order, neither of these statements is true. 

To avoid attachment to any particular definition of the conservative dynamics, I begin in Sec.~\ref{preview} with a preview of the main results, which hold for most, if not all, specifications of the conservative-dissipative split. Without making a precise choice of that split, I sketch the derivation of a general formula for the second-order $\btilde u^t$.

Sections~\ref{SC} and \ref{GW} then describe a particular definition of the conservative dynamics, eventually recovering the result for $\btilde u^t$. In Sec.~\ref{SC} I offer a description in the self-consistent self-force formalism~\cite{Pound:10a, Pound:12a, Pound:12b, Pound:13a, Pound:14a, Pound-Miller:14}, a picture of the system in which the metric perturbation is a functional of the self-accelerated orbit. After a review of the formalism, I construct a precisely circular orbit that is a geodesic of a certain time-symmetrized effective metric constructed from the retarded field, and I derive a gauge-invariant formula for the second-order $\btilde u^t$ on that orbit.

The self-consistent formalism is not ideal for numerical calculations of conservative dynamics, for reasons described below, and so in Sec.~\ref{GW} I transition to a Gralla-Wald picture, in which the perturbed motion is described as a small deviation from a reference orbit that is a geodesic of the background spacetime~\cite{Gralla-Wald:08,Gralla:12}. Although this description of the motion is not ideal for describing dissipative changes in the orbit, which grow large with time, it \emph{is} ideal for calculations of conservative dynamics, because in the absence of dissipation, deviations from the reference orbit remain small. Beginning from the self-consistent results of Sec.~\ref{SC}, I derive an expression for the second-order redshift variable in the Gralla-Wald picture. Section~\ref{gauge_GW} shows the gauge invariance of the result.

In Sec.~\ref{variants} I briefly discuss alternative definitions of the conservative dynamics. The formula for $\btilde u^t$ holds true with these definitions, but some difficulties arise in interpreting that formula and enforcing its gauge invariance.

I conclude in Sec.~\ref{scheme} by describing a numerical scheme for calculating $\btilde u^t$ in the frequency domain in the Gralla-Wald picture. The scheme is an extension of one recently devised by Warburton and Wardell for the scalar self-force problem~\cite{Warburton-Wardell:14}. Its technical details will be provided in a future paper~\cite{Warburton-etal:14}.

Appendix~\ref{dissipation} complements the body of the paper with a treatment of quasicircular orbits in the Gralla-Wald picture, relying less on the self-consistent picture.

I work in geometric units with $G=c=1$, and I use the metric signature $-+++$. All indices are raised and lowered with a background metric $g_{\mu\nu}$, both a semicolon and $\nabla$ denote the covariant derivative compatible with $g_{\mu\nu}$, and coordinate expressions always refer to Schwarzschild coordinates $\{t,r,\theta,\phi\}$ on the background manifold.

\section{Preview}\label{preview}
Consider a compact, slowly spinning, nearly spherical object of mass $m$ moving about a Schwarzschild black hole of mass $M\gg m$. If we split the binary's full metric ${\sf g}_{\mu\nu}$ into the Schwarzschild background $g_{\mu\nu}$ and a perturbation $h_{\mu\nu}\equiv {\sf g}_{\mu\nu}-g_{\mu\nu}$, then in the background, the object of mass $m$ moves on a worldline $z^\mu$ governed by the equation of motion~\cite{Pound:12a,Pound:14a}
\begin{equation}
\frac{D^2 z^\mu}{d\tau^2} = F^\mu,\label{SC_motion_prev}
\end{equation}
where $F^\mu$ is the self-force per unit mass, given by
\begin{equation}
F^\mu = -\frac{1}{2}P^{\mu\nu}(g_\nu{}^\delta-h^\R_\nu{}^\delta)(2h^\R_{\delta\beta;\gamma}-h^\R_{\beta\gamma;\delta})u^\beta u^\gamma+O(\e^3).\label{force}
\end{equation}
Here $\tau$ and $u^\mu \equiv \frac{dz^\mu}{d\tau}$ are the proper time and four-velocity along $z^\mu$ as normalized in $g_{\mu\nu}$, $P^{\mu\nu}\equiv g^{\mu\nu}+u^\mu u^\nu$ projects orthogonally to $u^\mu$, and $\e\equiv1$ is used to count powers of the mass ratio $m/M$. The key quantity appearing in the self-force is $h^\R_{\mu\nu}$, called the \emph{regular field}, a certain smooth vacuum perturbation made up of pieces of $h_{\mu\nu}$. The particular regular field appearing here is discussed in Sec.~\ref{SC_formalism} and defined precisely in Refs.~\cite{Pound:12a,Pound:12b,Pound-Miller:14}. It contains both first- and second-order contributions, and I write it as $h^\R_{\mu\nu}=\e h^{\R1}_{\mu\nu}+\e^2 h^{\R2}_{\mu\nu}$. In lieu of its precise definition, it can be thought of as a nonlinear generalization of the familiar Detweiler-Whiting regular field~\cite{Detweiler-Whiting:03}, given by Eq.~\eqref{hR1_Greens} below. 

The equation of motion can be cast in more compelling form if instead we write it in an \emph{effective metric} $\btilde g_{\mu\nu}\equiv g_{\mu\nu}+h^\R_{\mu\nu}$, which, by construction, is a $C^\infty$ solution to the vacuum Einstein equations. After reparametrizing the worldline with proper time $\btilde\tau$ measured in $\btilde g_{\mu\nu}$, and converting to the covariant derivative $\btilde \nabla_{\!\mu}$ compatible with $\btilde g_{\mu\nu}$, one finds that the equation of motion~\eqref{SC_motion_prev} in $g_{\mu\nu}$ becomes the geodesic equation in $\btilde g_{\mu\nu}$:
\begin{equation}\label{geodesic_form_prev}
\frac{\btilde D^2z^\mu}{d\btilde\tau^2}=O(\epsilon^3);
\end{equation} 
in other words, the object is in freefall in the vacuum field $\btilde g_{\mu\nu}$. Here  $\frac{\btilde D}{d\btilde\tau}\equiv \btilde u^\mu\btilde\nabla_{\!\mu}$, with $\btilde u^\mu\equiv \frac{dz^\mu}{d\btilde\tau}$. Although the statements in this and the preceding paragraph have been derived only in gauges smoothly related to the Lorenz gauge~\cite{Pound:12a,Pound:14a}, Detweiler has heuristically argued that they should be true in any sufficiently well-behaved gauge~\cite{Detweiler:12}.

\begin{figure}[t]
\begin{center}
\includegraphics[width=0.75\columnwidth]{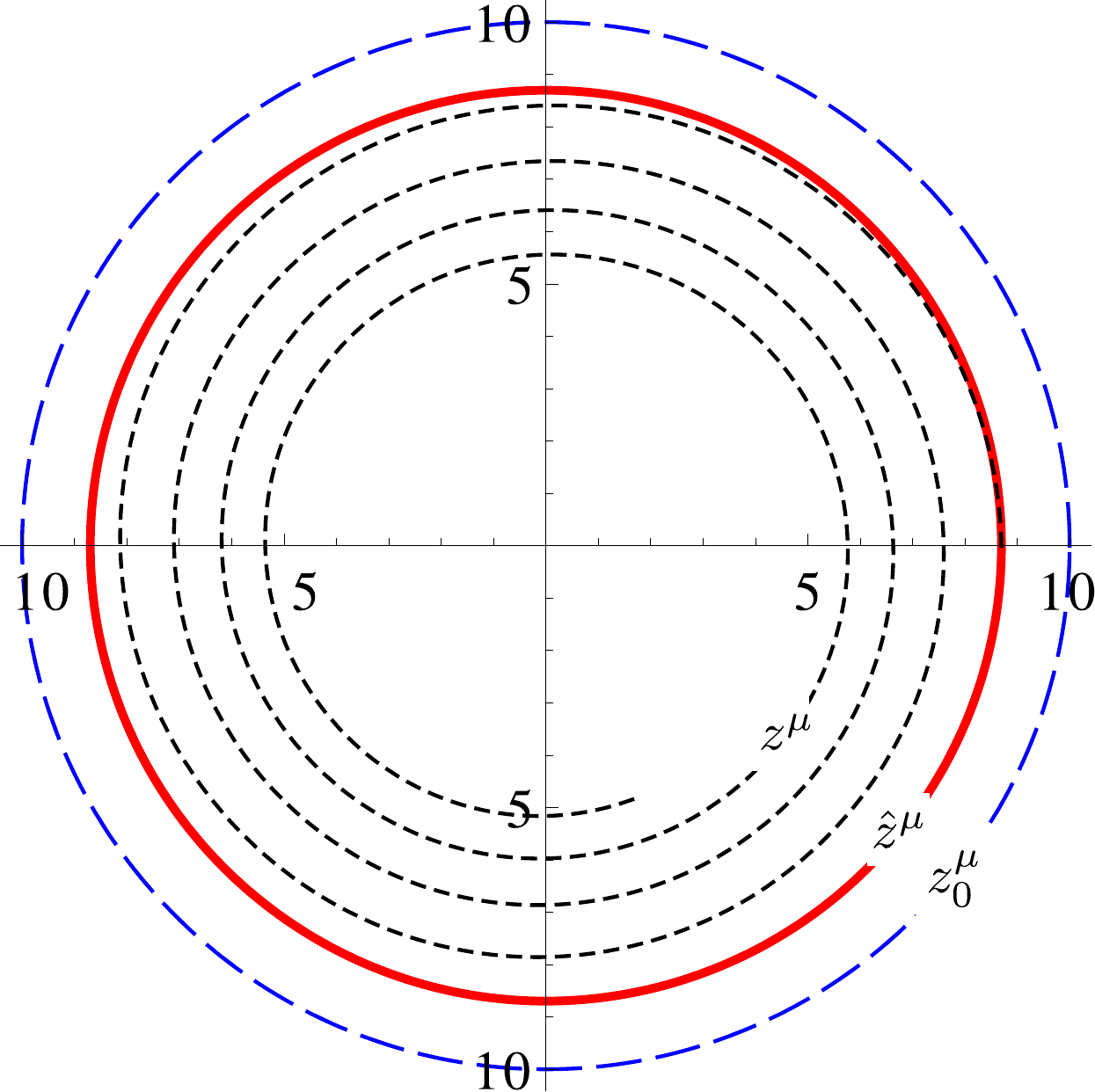}
\caption{\label{orbits}The inspiraling orbit $z^\mu$, conservatively accelerated orbit $\hat z^\mu$, and zeroth-order, background-geodesic orbit $z_0^\mu$ in the equatorial plane. Distances are displayed in units of $M$. The accelerated orbits $\hat z^\mu$ and $z^\mu$ are calculated from the first-order Gralla-Wald approximation in Eqs.~\eqref{r1_diss}--\eqref{phi1_diss}, using the Lorenz-gauge self-force library from Ref.~\cite{Warburton-etal:12}, a mass ratio of $M/m=3$, a zeroth-order orbital radius $r_0=10M$, and an initial angle $\phi(0)=0$. The zeroth-order orbit is chosen to have the same orbital frequency as $\hat z^\mu$.}
\end{center}
\end{figure}

Now consider the case of interest in this paper: take $z^\mu$ to be a quasicircular orbit, precisely circular but for dissipation. Suppose that in one way or another, I artificially ``turn off" the dissipative effect of the self-force, and let $\hat z^\mu$ denote the resulting conservative, circular orbit. The coordinate form of a `circular' orbit can be almost arbitrarily altered by a gauge transformation, under which $\hat z^\mu\to \hat z^\mu-\e \xi^\mu+O(\e^2)$; the gauge freedom in the formalism is discussed in Sec.~\ref{gauge_GW}. To maintain some degree of physical intuition, I restrict the discussion to `nice' gauges, in which the conservative orbit can be parametrized in the manifestly circular form
\begin{equation}\label{circ_prev}
\hat z^\mu(t,\epsilon) = \{t,\ \hat r(\epsilon),\ \pi/2,\ \Omega(\epsilon) t\},
\end{equation}
where I have placed the orbit on the equatorial plane and introduced the orbital frequency $\Omega\equiv \frac{d\hat\phi}{dt}$. This orbit must satisfy an equation of motion with a purely radial, constant force, call it $\hat F^\mu = \delta^\mu_r \hat F^r$, such that 
\begin{equation}\label{SC_con_motion_prev}
\frac{D^2\hat z^\mu}{d\tau^2} = \hat F^\mu. 
\end{equation}
The relationship between $z^\mu$ and $\hat z^\mu$ is shown schematically in Fig.~\ref{orbits}.

Following Detweiler~\cite{Blackburn-Detweiler:92,Detweiler:08}, I use the orbital frequency to define a helical vector
\begin{equation}\label{helical_prev}
k^\alpha = \{1,0,0,\Omega\}.
\end{equation}
As discussed in Sec.~\ref{symmetry-of-field}, the perturbed spacetime inherits the orbit's helical symmetry, and in a gauge compatible with Eq.~\eqref{circ_prev}, $k^\mu$ will be a Killing vector of the perturbed spacetime. The four-velocity on $\hat z^\mu$, 
\begin{equation}
\hat u^\alpha = \hat U(\epsilon)\{1,0,0,\Omega(\epsilon)\}= \hat Uk^\alpha,\label{u_prev}
\end{equation}
is parallel to it. The proportionality factor is $\hat U\equiv \frac{dt}{d\tau} = \hat u^t$, the ratio of coordinate time to proper time (as measured in $g_{\mu\nu}$) on $\hat z^\mu$. Of course, the four-velocity as normalized in the effective metric $\btilde g_{\mu\nu}$ is likewise parallel to $k^\mu$, 
\begin{equation}
\btilde u^\alpha = \btilde U(\epsilon)k^\mu, \label{utilde_prev}
\end{equation}
with a proportionality factor
\begin{equation}\label{Utilde_prev}
\btilde U \equiv \frac{dt}{d\btilde\tau} = \btilde u^t. 
\end{equation}
This last quantity (rather than its inverse) is what I will call Detweiler's redshift variable, the ratio of coordinate time to proper time as measured in $\btilde g_{\mu\nu}$ on $\hat z^\mu$.

A formula for $\btilde U$ can be found from the equation of motion~\eqref{SC_con_motion_prev} and the normalization conditions $\btilde g_{\mu\nu}\btilde u^\mu \btilde u^\nu = -1$ and $g_{\mu\nu}\hat u^\mu \hat u^\nu=-1$, together with Eqs.~\eqref{u_prev} and \eqref{utilde_prev} for the four-velocity. The few, simple steps involved in that calculation are shown in Sec.~\ref{U_SC}. Their end result is the following:
\begin{multline}
\btilde U = (1-3M/\hat r)^{-1/2}\left\{1+\frac{1}{2}(h^\R_{uu}-\hat F_r \hat r)\right.\\
\left.+\frac{1}{8}\left[3(h^\R_{uu})^2-2\hat r \hat F_{r} h^\R_{uu}-\hat r^2(\hat F_r)^2\right]+O(\epsilon^3)\right\},\label{Utilde_SC_prev}
\end{multline}
where $h^\R_{uu}\equiv h^\R_{\mu\nu}\hat u^\mu \hat u^\nu$, and I have used the fact that $h^\R_{uu}\sim \hat F_r\sim\e$.

Equation \eqref{Utilde_SC_prev} yields $\btilde U$ in the self-consistent picture. I discuss this picture in Sec.~\ref{SC_formalism}, but for now I merely state that in it, the field equations are coupled to the equation of motion~\eqref{SC_motion_prev}. The metric perturbation is generated by $m$'s self-forced motion, and the self-force is constructed from (the regular piece of) that same perturbation; the perturbation is \emph{not} sourced by geodesic motion on the background spacetime, as it often is in leading-order approximations in the self-force literature~\cite{Barack:09,Poisson-Pound-Vega:11}.  Therefore, the quantities in Eq.~\eqref{Utilde_SC_prev} are both evaluated on the orbit $\hat z^\mu$ and are constructed from fields sourced by that orbit. This self-consistent approach is an ideal way of going about things when including dissipation in the dynamics, because it correctly accounts for long-term, dissipative changes in the orbit~\cite{Pound:10a}. But it is impractical in the present case, because one must know the radial force (and therefore the regular field) in order to find the accelerated circular orbit $\hat z^\mu$, and at the same time one must know $\hat z^\mu$ in order to find the regular field (and therefore the radial force).

To simplify the problem, I transition from the self-consistent picture to a Gralla-Wald one. In the Gralla-Wald picture, the orbit is expanded in a Taylor series around a zeroth-order geodesic of the background spacetime, $z_0^\mu$. That is, in the present case,
\begin{equation}
\hat z^\mu(t,\epsilon) = z^\mu_0(t)+\epsilon\hat z^\mu_1(t)+\epsilon^2\hat z^\mu_2(t)+O(\epsilon^3);\label{expanded_zhat}
\end{equation}
no hat is required over $z^\mu_0$, because the zeroth-order term in $\hat z^\mu$ can be chosen to be identical to that in the inspiraling orbit $z^\mu$. If dissipation were accounted for, the corrections $\hat z_{n}^\mu$ in this expansion would quickly grow large, and the approximation would break down. But since only conservative effects are accounted for here, these corrections remain small, making the expansion of the orbit quite convenient. It allows us to freely specify the zeroth-order orbit and then proceed sequentially to the first-order field sourced by $z_0^\mu$, the correction to the motion $\hat z_1^\mu$ due to that perturbation, and so on; I refer the reader to Sec.~\ref{GW_formalism} for a more detailed description.
 
After deciding to expand $\hat z^\mu$ around a background geodesic, we are still left with the freedom to decide \emph{which} background geodesic to expand around. This freedom persists even after all the standard gauge freedom of perturbation theory is exhausted; it would exist even if we were considering the expansion of a perturbed orbit about a Keplerian one in fixed coordinates in Newtonian physics, for example. For my purposes here, the most convenient choice of reference geodesic is another circular orbit of the same orbital frequency $\Omega$ as $\hat z^\mu$. The zeroth-order worldline is then
\begin{equation}
z_0^\mu = \left\lbrace t, r_0, \frac{\pi}{2}, \Omega t\right\rbrace,\label{z0_prev}
\end{equation}
with the relationship between orbital frequency and radius given by the familiar geodesic formula
\begin{equation}
\Omega = \sqrt{\frac{M}{r_0^3}}.\label{Omega0_prev}
\end{equation}
Again, the relationship between this orbit and the perturbed ones is displayed schematically in Fig.~\eqref{orbits}; because $\hat z^\mu$ and $z_0^\mu$ share the same frequency, they differ only by radial corrections. The four-velocity on $z_0^\mu$ is again parallel to the helical Killing vector,  
\begin{equation}
u^\mu_0\equiv \frac{dz_0^\mu}{d\tau_0}=U_0 k^\alpha, 
\end{equation}
where $\tau_0$ is the proper time (measured in $g_{\mu\nu}$) on $z_0^\mu$, and $U_0\equiv \frac{dt}{d\tau_0}=1/\sqrt{1-3M/r_0}$. 

To utilize this expansion of the orbit, we must account for the fact that the fields we began with in the self-consistent picture depended both on the point $x^\mu$ at which they were evaluated and the source orbit $\hat z^\mu$ that produced them. For example, we can write the terms in the regular field as $h^{\R n}_{\mu\nu}(x;\hat z)$. When evaluating them at a point on $\hat z^\mu$, they read $h^{\R n}_{\mu\nu}(\hat z;\hat z)$, and both the first and second argument must be expanded. Section~\ref{GW_formalism} describes that procedure, and Secs.~\ref{expanded_motion} and \ref{Utilde_GW} provide the details of the expansion of $\btilde U$. The end result is
\begin{multline}
\btilde U = U_0\left\{1+\frac{1}{2}\epsilon  h^{\R1}_{u_0u_0}+\epsilon^2\left[\frac{1}{2} h^{\R2}_{u_0u_0}+\frac{3}{8}( h^{\R1}_{u_0u_0})^2\right.\right.\\
\left.\left.-\frac{r_0^2}{6M}(r_0-3M)(\hat F_{1r})^2\right]+O(\epsilon^3)\right\}\label{Utilde_GW_prev},
\end{multline}
where $h^{\R n}_{u_0u_0}\equiv h^{\R n}_{\mu\nu}u_0^\mu u_0^\nu$. $h^{\R1}_{\mu\nu}$ is now the usual linearized regular field produced by a point particle moving on $z_0^\mu$, rather than on $\hat z^\mu$, and $h^{\R2}_{\mu\nu}$ now incorporates the effect of translating the source worldline by an amount $\hat z^\mu_1$, in a manner described in Sec.~\ref{GW_formalism}. Regardless of the definition of conservative dynamics, it follows from Eq.~\eqref{SC_motion_prev} that the radial force (with index down) is given by $\hat F_{1r} = \frac{1}{2}h^{\R1}_{u_0u_0,r}$. 

Equation~\eqref{Utilde_GW_prev} is the main result of this paper. It describes the gauge-invariant ratio $dt/d\btilde\tau$ along the circular orbit $\hat z^\mu$, but each quantity in the formula is calculated on the zeroth-order worldline $z_0^\mu$, not on $\hat z^\mu$. In the following sections, I will provide all the details of its derivation, and in Sec.~\ref{gauge_GW} I will explicitly show its invariance. Moreover, I will describe different formulations of conservative dynamics that lead to it, and its precise interpretation in each case. Along the way, I will describe most of the tools necessary to concretely calculate the quantities $h^{\R1}_{u_0u_0}$, $h^{\R 2}_{u_0u_0}$, and $\hat F_{1r}$ that appear in the formula.

Before moving onto that discussion, I note that Eq.~\eqref{Utilde_GW_prev} is not yet in a suitable form for comparison with other models, such as PN theory. Although it is gauge independent in the sense of perturbation theory, it still depends on the Schwarzschild coordinate radius of $z^\mu_0$. To put it in a coordinate-independent form, I use Eq.~\eqref{Omega0_prev} to replace $r_0$ with $(M/\Omega^2)^{1/3}$. The result is
\begin{equation}
\btilde U = U_0(\Omega) + \e\btilde U_1(\Omega) + \e^2\btilde U_2(\Omega) +O(\e^3),\label{UvsOmega}
\end{equation}
where
\begin{align}
U_0(\Omega) &= \frac{1}{\sqrt{1-3(M\Omega)^{2/3}}}, \\
\btilde U_1(\Omega) &= \frac{1}{2}U_0(\Omega)h^{\R1}_{u_0u_0}, \\
\btilde U_2(\Omega) &= U_0(\Omega)\bigg\{\frac{1}{2}h^{\R2}_{u_0u_0}+\frac{3}{8}\left(h^{\R1}_{u_0u_0}\right)^2\nonumber\\
						&\qquad\qquad	-\frac{1}{6\Omega^2}\left(F_{1r}\right)^2 \left[1-3 (M\Omega)^{2/3}\right]\bigg\}.
\end{align}
$U_2(\Omega)$ is the new term not previously calculated in a self-force formalism. In PN theory, $\btilde U$ is often written in a different coordinate-independent way, as a function of a variable $x\equiv [(M+m)\Omega]^{2/3}$. One can easily do the same here by using the expansion $M\Omega=x^{3/2}\left[1-\e\frac{m}{M}+\e^2\left(\frac{m}{M}\right)^2+O(\e^3)\right]$, but for the sake of brevity I omit the resulting (lengthier) expression for $\btilde U(x)$. 




\section{Self-consistent picture: an accelerated worldline}\label{SC}

\subsection{Formalism}\label{SC_formalism}
To place the above preview in proper context, and to lead up to the definitions of conservative dynamics, I now review the second-order self-force formalism. In general terms, all self-force formalisms are designed to model the perturbation produced by a small object \emph{without} modeling the details of the object's internal structure. But each formalism achieves this in a slightly different way. Here I use the self-consistent approximation scheme presented in Ref.~\cite{Pound:10a} and further developed in Refs.~\cite{Pound:12a,Pound:12b,Pound:14a}. I refer the reader to the reviews~\cite{Poisson-Pound-Vega:11,Barack:09} for a broader description of self-forces in curved spacetimes and a pedagogical introduction to many of the technical tools used in the field. 

The metric perturbation in the self-consistent scheme is written as an expansion
\begin{equation}
h_{\mu\nu} = \epsilon h^1_{\mu\nu}[z] + \epsilon^2 h^2_{\mu\nu}[z]+O(\epsilon^3),
\end{equation}
where $\epsilon\equiv1$ is used to count powers of $m/M$, and each term in the expansion is a functional of the self-accelerated worldline $z^\mu$, which represents, in a rough sense, the small object's center of mass.\footnote{If the object is a black hole, clearly there is no timelike worldline in its interior that represents its ``center of mass". Nevertheless, $z^\mu$ can be interpreted that way even for black holes, exotic objects containing worm holes, etc.; see Refs.~\cite{Pound:10a,Pound:12b,Pound:14a} for the precise definition of $z^\mu$.}  The Lorenz gauge condition $\nabla^\mu\bar h_{\mu\nu}=0$, where an overbar indicates trace-reversal, is imposed on the total perturbation but not on any individual term $h^n_{\mu\nu}$. Section~\ref{gauge_GW} briefly describes the transformation to other gauges. More detailed descriptions of the formalism's gauge freedom will be presented in Refs.~\cite{Pound:14a,Pound:14b}.

To disregard unneeded information about the object's internal structure, one examines the general solution to the Einstein equation in a small vacuum region \emph{outside} the object; there, the metric depends on the object's composition only through bulk variables such as mass and spin. The equation of motion governing $z^\mu$ follows from imposing an appropriate centeredness condition on the metric in this region. For a sufficiently spherical and slowly spinning object, the result through second order is Eq.~\eqref{SC_motion_prev}. Since the regular field inherits $h_{\mu\nu}$'s functional dependence on the worldline, I write it here as
\begin{equation}
h^\R_{\mu\nu} = \epsilon h^{\R1}_{\mu\nu}[z]+\epsilon^2 h^{\R2}_{\mu\nu}[z] + O(\epsilon^3).
\end{equation}
Of course, one can always extract different smooth pieces from any metric. The \emph{particular} regular field I use here, the one that appears in the equation of motion, is described in Refs.~\cite{Pound:10a,Pound:12a,Pound:12b,Pound-Miller:14}. It is defined such that its value (and those of its derivatives) on $z^\mu$ are equal to certain pieces of $h_{\mu\nu}$ in the object's exterior; this is how an analysis of the field outside the object yields an equation of motion in terms of variables on a worldline effectively inside the object. Although the precise definition of the regular field is somewhat technical, involving a decomposition of the metric into harmonics around the object, one can think of $h^{\R}_{\mu\nu}$ informally as the piece of $h_{\mu\nu}$ that does not depend on local information about the object. As implied in the preview, for the purposes of this paper, the first-order term, $h^{\R1}_{\mu\nu}$,  in the Lorenz gauge can be taken to be the well-known Detweiler-Whiting regular field~\cite{Detweiler-Whiting:03}. A suitable definition of $h^\R_{\mu\nu}$ in other gauges is given in Sec.~\ref{gauge_GW} below.

The value of the regular field, since it is not determined by local information, must be found by solving the Einstein equation globally. In general, this means solving the equation numerically, which can be achieved using a \emph{puncture scheme}, as has been done at first order~\cite{Barack-Golbourn:07,Vega-Detweiler:07,Wardell-etal:11}. First, the field $h_{\mu\nu}$ found outside the object is analytically continued into its interior, and a \emph{singular field} $h^\S_{\mu\nu}\equiv h_{\mu\nu}-h^\R_{\mu\nu}$ is defined. This field diverges on $z^\mu$, behaving (schematically) as
\begin{align}
h^{\S 1}_{\mu\nu}&\sim \frac{m}{|x^\alpha-z^\alpha|} + O(|x^\alpha-z^\alpha|^0),\label{hS1_SC_schematic}\\
h^{\S 2}_{\mu\nu}&\sim \frac{m^2}{|x^\alpha-z^\alpha|^2} + \frac{\delta m_{\mu\nu}+mh^{\R1}_{\mu\nu}}{|x^\alpha-z^\alpha|} 
			+ O(\ln |x^\alpha-z^\alpha|),\label{hS2_SC_schematic}
\end{align}
where $|x^\alpha-z^\alpha|$ represents a measure of spatial distance from $z^\mu$. In the second term in Eq.~\eqref{hS2_SC_schematic}, $h^{\R1}_{\mu\nu}$ is evaluated on $z^\mu$; at higher orders in $|x^\alpha-z^\alpha|$, derivatives of $h^{\R1}_{\mu\nu}$ on $z^\mu$ appear. Also in that second term is the quantity
\begin{align}
\delta m_{\alpha\beta} &= m(g_{\alpha\beta}+2u_{\alpha} u_{\beta})u^\mu u^\nu h^{\R1}_{\mu\nu}\nonumber\\
					&\quad +\frac{1}{3}m\left(2h^{\R1}_{\alpha\beta}+g_{\alpha\beta}g^{\mu\nu}h^{\R1}_{\mu\nu}\right)
						+4mu_{(\alpha}h^{\R1}_{\beta)\mu}u^\mu,
\end{align}
a tensor on $z^\mu$ that can be interpreted as a gravitational correction to the object's monopole moment. Here I have presented the local expansions of $h^{\S n}_{\mu\nu}$ only schematically, but they can be found in explicit, covariant form in Ref.~\cite{Pound-Miller:14}. From those local expansions, one can construct \emph{punctures} that capture the irregularity in the (analytically continued) physical field $h_{\mu\nu}$, and one can then replace the field equations for $h_{\mu\nu}$ with field equations for the regular part of $h_{\mu\nu}$, thereby replacing the physical system with an effective one. 

This is done as follows: choose punctures $h^{\P n}_{\mu\nu}$, which can be any fields that locally approximate $h^{\S n}_{\mu\nu}$ near $z^\mu$, and define \emph{residual fields} $h^{\res n}_{\mu\nu}\equiv h^{n}_{\mu\nu}-h^{\P n}_{\mu\nu}\approx h^{\R n}_{\mu\nu}$. If $h^{\P n}_{\mu\nu}$ is a good enough local approximation to $h^{\S n}_{\mu\nu}$, then we will have $h^{\res n}_{\mu\nu}\big|_z=h^{\R n}_{\mu\nu}\big|_z$ and $h^{\res n}_{\mu\nu;\sigma}\big|_z=h^{\R n}_{\mu\nu;\sigma}\big|_z$ \emph{exactly}, even though off the worldline $h^{\res n}_{\mu\nu}$ will only approximate $h^{\R n}_{\mu\nu}$. Letting $\Gamma$ be a worldtube enclosing the object, we may solve for the effective, residual fields inside $\Gamma$ and for the physical fields outside. The puncture scheme (in the Lorenz gauge) is then encapsulated by the field equations\footnote{These equations are generally written with distributional stress-energies on their right-hand sides, which cancel distributional content in $E_{\mu\nu}[h^{\P n}]$. Here I follow Gralla~\cite{Gralla:12} in considering the sources on the right-hand side to be defined only off $z^\mu$, which suffices to uniquely determine the solutions both off \emph{and on} $z^\mu$.}
\begin{subequations}\label{h1_SC}%
\begin{align}
E_{\mu\nu}[h^{\res1}] &= -E_{\mu\nu}[h^{\P1}] \equiv S^{1\rm eff}_{\mu\nu}& \text{inside }\Gamma,\label{h1_SC_in}\\
E_{\mu\nu}[h^{1}] &= 0 & \text{outside }\Gamma,
\end{align}
\vspace{-1.5\baselineskip}
\end{subequations}
\begin{subequations}\label{h2_SC}%
\begin{align}
E_{\mu\nu}[h^{\res2}] &= 2\delta^2R_{\mu\nu}[h^1,h^1]- E_{\mu\nu}[h^{\P2}]\hspace{-10pt} &\nonumber\\
											& \equiv S^{2\rm eff}_{\mu\nu}  & \text{inside }\Gamma,\label{h2_SC_in}\\
E_{\mu\nu}[h^2] &= 2\delta^2R_{\mu\nu}[h^1,h^1] & \text{outside }\Gamma,\label{h2_SC_out}
\end{align}
\end{subequations}
where $E_{\mu\nu}[h]\equiv \Box h_{\mu\nu}+2R_\mu{}^\alpha{}_\nu{}^\beta h_{\alpha\beta}$ is the usual tensorial wave operator, and
\begin{align}
\delta^2R_{\alpha\beta}[h,h] &=-\tfrac{1}{2}h^{\mu\nu}\left(2h_{\mu(\alpha;\beta)\nu}-h_{\alpha\beta;\mu\nu}-h_{\mu\nu;\alpha\beta}\right)
						\nonumber\\
					&\quad +\tfrac{1}{4}h^{\mu\nu}{}_{;\alpha}h_{\mu\nu;\beta}
					+\tfrac{1}{2}h^{\mu}{}_{\beta}{}^{;\nu}\left(h_{\mu\alpha;\nu} -h_{\nu\alpha;\mu}\right)\nonumber\\
					&\quad -\tfrac{1}{2}\bar h^{\mu\nu}{}_{;\nu}\left(2h_{\mu(\alpha;\beta)}-h_{\alpha\beta;\mu}\right) \label{d2R}
\end{align}
is the quadratic term in the expansion of the Ricci tensor $R_{\mu\nu}[g+h]$. Both $\delta^2R_{\alpha\beta}[h^1,h^1]$ and $E_{\mu\nu}[h^{\P2}]$ diverge as $1/|x^\alpha-z^\alpha|^4$ near the worldline; this can be seen schematically from Eq.~\eqref{hS2_SC_schematic}. But as a consequence of the puncture's construction, in Eq.~\eqref{h2_SC_in} the divergence of $E_{\mu\nu}[h^{\P2}]$ cancels that of $\delta^2R_{\alpha\beta}[h^1,h^1]$ to leave a source $S^{2\rm eff}_{\mu\nu}$ that is sufficiently regular to obtain a well-defined solution. Similarly,  in Eq.~\eqref{h1_SC_in} $E_{\mu\nu}[h^{\P1}]$ will contain terms that diverge as $1/|x^\alpha-z^\alpha|^3$, which can be seen from Eq.~\eqref{hS1_SC_schematic}, but these terms cancel to leave an integrable source $S^{1\rm eff}_{\mu\nu}$. The better the punctures $h^{\P n}_{\mu\nu}$ approximate $h^{\S n}_{\mu\nu}$, the better $h^{\res n}_{\mu\nu}$ approximates $h^{\R n}_{\mu\nu}$, and the closer the field equations inside $\Gamma$ get to the vacuum equations $E_{\mu\nu}[h^{\R 1}]=0$ and $E_{\mu\nu}[h^{\R2}]=2\delta^2R_{\mu\nu}[h^{\R1},h^{\R1}]$.

The field equations~\eqref{h1_SC} and \eqref{h2_SC} on their own are incomplete, because they require one to know the trajectory of the puncture. The complete system is composed of the field equations coupled to the equation of motion~\eqref{SC_motion_prev}, with $h^{\R n}_{\mu\nu}$ replaced by $h^{\res n}_{\mu\nu}$ in Eq.~\eqref{force}; this dependence on the puncture's motion implicitly defines the functionals $h^n_{\mu\nu}[z]$ and $h^{\R n}_{\mu\nu}[z]$. By solving the coupled system, one self-consistently determines the orbit and the fields. 


To relate this discussion to typical treatments of the first-order problem, I note that the field $h^1_{\mu\nu}$ found outside the object is identical to one sourced by a point mass moving on $z^\mu$: its analytical continuation to $z^\mu$ satisfies
\begin{align}
E_{\mu\nu}[\bar h^1] &= -16\pi \int_z mu_\mu u_\nu \frac{\delta^4(x^\alpha-z^\alpha)}{\sqrt{-g}}d\tau\label{h1_SC_point}\\
										&\equiv -16\pi T^1_{\mu\nu}[z],\label{T1}
\end{align}
where $g$ is the determinant of $g_{\mu\nu}$. This means that at linear order, the correct physical solution outside the object can be obtained by modeling the object as a point mass. Proofs of this statement can be found in Refs.~\cite{DEath:75,Gralla-Wald:08,Pound:10a,Pound:12b}. From this perspective, to obtain the regular field one could solve Eq.~\eqref{h1_SC_point} and then subtract the singular field from the result. 
 Alternatively, the regular field can be written as an explicit, rather than implicit, functional of $z^\mu$ using the Detweiler-Whiting regular Green's function~\cite{Detweiler-Whiting:03}:
\begin{equation}
h^{\R1}_{\mu\nu}[z] = 4m\int_z \bar G^\R_{\mu\nu\alpha'\beta'}(x,z(\tau))u^{\alpha'} u^{\beta'} d\tau,\label{hR1_Greens}
\end{equation}
where primed indices refer to tensors at $x'^\mu=z^\mu(\tau)$, and the overbar again indicates trace-reversal. 

At second order, the terms involving $\delta m_{\mu\nu}$ in $h^{\S2}_{\mu\nu}$, call them $h^{\delta m}_{\mu\nu}$, satisfy the analogous point-particle equation
\begin{align}
E_{\mu\nu}[\bar h^{\delta m}] &= -4\pi \int_z \overline{\delta m}_{\mu\nu} \frac{\delta^4(x^\alpha-z^\alpha)}{\sqrt{-g}}d\tau\label{hdm_point}\\
										&\equiv -16\pi T^2_{\mu\nu}[ z].
\end{align}
However, the remainder of the field $h^2_{\mu\nu}$ cannot be written as the solution to a distributional equation in this manner, and no equivalent to Eq.~\eqref{hR1_Greens} for the regular field $h^{\R2}_{\mu\nu}$ is yet known. At all points off $z^\mu$, $h^2_{\mu\nu}$ satisfies Eq.~\eqref{h2_SC_out}, but it does not satisfy a distributionally well-defined equation on any domain including $z^\mu$. This is a consequence of the fact that point-particle distributions cease to be useful models beyond linearized theory. With such sources, the nonlinear equations have no solution in any well-behaved space of functions; see Ref.~\cite{Steinbauer-Vickers:08} for a recent discussion. Hence, the problem must be tackled via an effective, regular field equation such as~\eqref{h2_SC}.

\subsection{Conservative dynamics}\label{dynamics_SC}
We are interested in the retarded solution to the coupled system made up of Eqs.~\eqref{SC_motion_prev}, \eqref{h1_SC}, and \eqref{h2_SC}. From this solution, I wish to extract the conservative dynamics. I now set about doing that.

In the coupled system, the retarded solution is represented by a triplet $(z^\mu,h^\R_{\mu\nu},h_{\mu\nu})$. My goal is to construct a certain ``subsystem,'' denoted by $(\hat z^\mu,\hat h^\R_{\mu\nu})$, that is purely conservative. The pair $(\hat z^\mu,\hat h^\R_{\mu\nu})$ will be such that $\hat z^\mu$ is precisely circular, $\hat h^\R_{\mu\nu}$ is time symmetric in an appropriate sense, and $\hat z^\mu$ is a geodesic of the effective metric  $\btilde g_{\mu\nu}=g_{\mu\nu}+\hat h^\R_{\mu\nu}$. This construction allows me to define a redshift variable $\btilde u^t$ by normalizing the four-velocity in the same metric in which the orbit is geodesic. Later, in Sec.~\ref{variants}, I will describe a construction that uses $h^\R_{\mu\nu}$ instead of $\hat h^\R_{\mu\nu}$.%

I first consider the consequences of replacing the quasicircular orbit $z^\mu$ with a precisely circular orbit $\hat z^\mu$; this can be thought of heuristically as ``turning off'' dissipation, although the ambiguity in that phrase will become clear below. After working out the broad features of the retarded field corresponding to a puncture moving on $\hat z^\mu$, I then extract a time-symmetrized effective metric from the retarded field and specify $\hat z^\mu$ to be a geodesic of that metric.

\subsubsection{Retarded field with a circular source}\label{symmetry-of-field}
There is considerable gauge freedom within the Lorenz gauge, meaning the conservative orbit can take multiple coordinate forms. I assume the particular gauge used is `nice', in the sense that the circular orbit $\hat z^\mu$ can be parametrized in the manifestly circular form~\eqref{circ_prev}. The four-velocity $\hat u^\mu$ is then given by $\hat u^\mu=\hat U k^\mu$, as previewed in Eq.~\eqref{u_prev}, with $\hat U\equiv \frac{dt}{d\tau} = \hat u^t$.

To study the retarded field corresponding to this orbit,\footnote{For simplicity, I assume the retarded field in the Lorenz gauge is unique, with no possibility of alteration by gauge modes. That is, I assume the equation $E_{\mu\nu}[h^n]=S^n_{\mu\nu}$ has a unique retarded solution for each source $S^n_{\mu\nu}$, although I am unaware of a proof of that proposition in Schwarzschild.} I leave the functionals $h^n_{\mu\nu}[z]$ and $h^{\R n}_{\mu\nu}[z]$ unchanged, simply replacing $z^\mu$ with $\hat z^\mu$. That is, the fields satisfy the puncture scheme composed of Eqs.~\eqref{h1_SC} and \eqref{h2_SC}, with the puncture moving on $\hat z^\mu$ instead of $z^\mu$. The entire system then inherits the orbit's helical symmetry. In other words, the metric perturbations satisfy the Killing equations
\begin{equation}
\mathcal{L}_k h^1_{\mu\nu}[\hat z] = 0, \qquad \mathcal{L}_k h^2_{\mu\nu}[\hat z] = 0,\label{helical_symmetry}
\end{equation}
and likewise for $h^{\R n}_{\mu\nu}$ and $h^{\S n}_{\mu\nu}$. On $\hat z^\mu$, these equations can be written as
\begin{equation}
\hat u^\rho h^{\R1}_{\mu\nu,\rho} = 0, \qquad \hat u^\rho h^{\R2}_{\mu\nu,\rho} = 0.\label{hR_const}
\end{equation}

These symmetries can be established concretely from that of the orbit. The source of the first-order equation in the form~\eqref{h1_SC_point}, evaluated in Schwarzschild coordinates, reads
\begin{equation}
T^1_{\mu\nu}[\hat z] = \frac{m\hat u_\mu \hat u_\nu}{ \hat r^2\hat U}\delta(r-\hat r)\delta(\theta-\pi/2)\delta(\phi-\Omega t), 
\end{equation}
which can be decomposed into ordinary scalar spherical harmonics as
\begin{equation}
T^1_{\mu\nu}[\hat z] = \frac{m\hat u_\mu \hat u_\nu}{\hat r^2\hat U}\delta(r-\hat r)\sum_{\ell m}Y^*_{\ell m}(\pi/2,\Omega t)Y_{\ell m}(\theta^A),\label{T1_Ylm}
\end{equation}
where $\theta^A=(\theta,\phi)$. This source has a time dependence $e^{-im\Omega t}$, and from its form one can infer that the retarded solution $h_{\mu\nu}^{1}$ has an expansion
\begin{equation}
h_{\mu\nu}^{1}(t,r,\theta^A;\hat z)=\sum_{i\ell m}h_{1i\ell m}(r;\hat r)e^{-im\Omega t}Y^{i\ell m}_{\mu\nu}(r,\theta^A),\label{h1_decomposition}
\end{equation}
where $h_{1i\ell m}$ satisfies the outgoing wave condition $h_{1i\ell m}\sim \frac{e^{ikr^*}}{r}$ at large $r$ and the ingoing wave condition $h_{1i\ell m}\sim e^{-ikr^*}$ at the horizon; here $r^*$ is the tortoise coordinate. As in Sec.~\ref{preview}, variables before a semicolon indicate the point at which the field is evaluated, while those after it indicate dependence on the source orbit. $Y^{i\ell m}_{\mu\nu}$ are the tensor spherical harmonics defined by Barack and Lousto~\cite{Barack-Lousto:05}, but any choice of tensor spherical harmonics would do. Each of the harmonics depends on $\phi$ only through an exponential $e^{im\phi}$, and to bring out the form of $h^1_{\mu\nu}$, I use that fact to rewrite Eq.~\eqref{h1_decomposition} as
\begin{equation}
h_{\mu\nu}^{1}(t,r,\theta^A;\hat z)=\sum_{i\ell m}H_{1i\ell m}(r;\hat r)e^{im(\phi-\Omega t)}P^{i\ell m}_{\mu\nu}(\theta),\label{h1_helical}
\end{equation}
with some appropriate functions $H_{1i\ell m}$ and $P^{i\ell m}_{\mu\nu}$. 
 In the form~\eqref{h1_helical}, $h^1_{\mu\nu}$ is manifestly helically symmetric. Naturally, $h^{\S1}_{\mu\nu}$ and $h^{\R1}_{\mu\nu}$ each possess this symmetry, and so $h^{\R1}_{\mu\nu}= constant$ on the worldline $\hat z^\mu$, where $\phi=\Omega t$. 

Similar considerations imply the helical symmetry of $h^2_{\mu\nu}[\hat z]$. We need only establish the symmetry of $T^2_{\mu\nu}[\hat z]$ and $\delta^2 R_{\mu\nu}$. The decomposition of $T^2_{\mu\nu}[\hat z]$ is essentially identical to that of $T^1_{\mu\nu}$, so I focus on $\delta^2 R_{\mu\nu}[h^1,h^1]$. By substituting the decomposition of $h^1_{\mu\nu}$ from Eq.~\eqref{h1_decomposition} into Eq.~\eqref{d2R}, we can see that $\delta^2 R_{\mu\nu}[h^1,h^1]$ has the form of a sum over helically symmetric terms of the form $e^{i(m'+m'')(\phi-\Omega t)}$. In fact, $\delta^2 R_{\mu\nu}[h^1,h^1]$ has a harmonic expansion
\begin{multline}
\delta^2 R_{\mu\nu}[h^1,h^1] = \sum_{i\ell m}\delta^2 R_{i\ell m}(r;\hat r)e^{-im\Omega t}Y^{i\ell m}_{\mu\nu}(r,\theta^A)
\end{multline} 
with radial functions given by a coupling formula of the form
\begin{equation}
\delta^2 R_{i\ell m} = \sum_{\substack{i'\ell' m'\\i''\ell''m''}}\mathcal{D}^{i'\ell'm'i''\ell''m''}_{i\ell m}\left[h_{1i'\ell'm'},h_{1i''\ell''m''}\right],\label{coupling}
\end{equation}
where $\mathcal{D}^{i'\ell'm'i''\ell''m''}_{i\ell m}$ is a bilinear differential operator. The explicit, lengthy expressions in this coupling formula will be given in a future publication~\cite{Warburton-etal:14}. Based on the helical symmetry of its source, $h_{\mu\nu}^{2}$ can be expanded as
\begin{equation}
h_{\mu\nu}^{2}(t,r,\theta^A;\hat z)=\sum_{i\ell m}h_{2i\ell m}(r;\hat r)e^{-im\Omega t}Y^{i\ell m}_{\mu\nu}(r,\theta^A)\label{h2_decomposition}
\end{equation}
and put in the manifestly helically symmetric form
\begin{equation}
h_{\mu\nu}^{2}(t,r,\theta^A;\hat z)=\sum_{i\ell m}H_{2i\ell m}(r;\hat r)e^{im(\phi-\Omega t)}P^{i\ell m}_{\mu\nu}(\theta),\label{h2_helical}
\end{equation}
and likewise for $h^{\S2}_{\mu\nu}$ and $h^{\R2}_{\mu\nu}$.

\subsubsection{Time-symmetrized effective metric}\label{time-symmetrized_metric}
At this point I still have not specified the equation of motion determining $\hat z^\mu$; I have merely stated that the orbit is circular. Because I have neglected all the dissipative forces in Eq.~\eqref{SC_motion_prev}, clearly $\hat z^\mu$ cannot satisfy the geodesic equation~\eqref{geodesic_form_prev} in the effective metric $g_{\mu\nu}+h^\R_{\mu\nu}[\hat z]$, which will include dissipative terms. I now construct an effective metric $\btilde g_{\mu\nu}[\hat z]=g_{\mu\nu}+\hat h^\R_{\mu\nu}[\hat z]$ in which $\hat z^\mu$ \emph{can} be made a geodesic. 

If second-order effects are ignored, the conservative piece of Eq.~\eqref{force} is uniquely defined by constructing the force from a half-retarded-plus-half-advanced metric perturbation, and the orbit is a geodesic of the effective metric corresponding to that perturbation. Taking this as my inspiration, I follow an analogous procedure to define $\hat h^\R_{\mu\nu}$.

Let $h^{1}_{\mu\nu}[\hat z]\equiv h^{1{\rm ret}}_{\mu\nu}[\hat z]$ be the retarded solution to Eq.~\eqref{h1_SC_point} with source $T^1_{\mu\nu}[\hat z]$, and let $h^{\rm adv}_{\mu\nu}[\hat z]$ be the advanced solution. The harmonic modes of these two solutions are related in a simple way. Referring to the form~\eqref{h1_decomposition}, I note that once $e^{-im\Omega t}Y^{i\ell m}_{\mu\nu}$ has been factored out of Eq.~\eqref{h1_SC_point}, the radial functions $h^{\rm ret/adv}_{1i\ell m}(r)$ satisfy a linear differential equation with real coefficients and a real source. The difference between the two solutions is produced solely by a complex conjugation of the boundary conditions: the retarded solution satisfies the outgoing wave condition $h_{1i\ell m}\propto e^{ikr^*}$ at infinity and the ingoing wave condition $h_{1i\ell m}\propto e^{-ikr^*}$ at the horizon, while the advanced solution satisfies the complex conjugate of these conditions. It follows that the modes of the two solutions are related by\footnote{This argument is due to Leor Barack.}
\begin{equation}
h^{\rm adv}_{1i\ell m} = h^{{\rm ret}*}_{1i\ell m},
\end{equation}
where the asterisk denotes complex conjugation. Therefore the radial coefficients in the half-retarded-plus-half-advanced solution, $\hat h^{1}_{\mu\nu}[\hat z]=\frac{1}{2}h^{1{\rm ret}}_{\mu\nu}[\hat z]+\frac{1}{2}h^{1{\rm adv}}_{\mu\nu}[\hat z]$, are given by $\hat h_{1i\ell m} = \frac{1}{2}(h_{1i\ell m}+ h^{*}_{1i\ell m})$. Here I am interested not in this global field, but in an effective metric in a neighbourhood of the worldline. Hence, corresponding to the half-retarded-plus-half-advanced field I introduce a regular field $\hat h^{\R1}_{\mu\nu}=\sum_{i\ell m}\hat h^\R_{1i\ell m}e^{-im\Omega t}Y^{i\ell m}_{\mu\nu}$ with radial coefficients
\begin{equation}
\hat h^\R_{1i\ell m} \equiv \frac{1}{2}(h^\R_{1i\ell m}+ h^{\R *}_{1i\ell m}).
\end{equation}

Now I do the same for the regular field at second order. I consider the retarded solution to Eq.~\eqref{h2_SC}, with $\delta^2 R_{\mu\nu}[h^1,h^1]$ constructed from the first-order retarded field, and with the second-order singular field that involves $h^{\R1}_{\mu\nu}$ in Eq.~\eqref{hS2_SC_schematic}, \emph{not} $\hat h^{\R1}_{\mu\nu}$. From the regular field $h^{\R2}_{\mu\nu}$ in this solution, I define a time-symmetrized regular field $\hat h^{\R2}_{\mu\nu}$ with radial coefficients
\begin{equation}
\hat h^\R_{2i\ell m} \equiv \frac{1}{2}(h^\R_{2i\ell m}+ h^{\R *}_{2i\ell m}).
\end{equation} 
This can be loosely thought of as the regular field corresponding to the half-retarded-plus-half-advanced solution to Eq.~\eqref{h2_SC}, but for reasons I discuss in Sec.~\ref{variants}, it is unlikely that such a solution would be globally well behaved. 

The time-symmetrized regular fields $\hat h^{\R n}_{\mu\nu}$ together define an effective metric $\btilde g_{\mu\nu}=g_{\mu\nu}+\hat h^\R_{\mu\nu}$, with 
\begin{equation}
\hat h^\R_{\mu\nu} \equiv \e \hat h^{\R1}_{\mu\nu}[\hat z]+\e^2\hat h^{\R2}_{\mu\nu}[\hat z].
\end{equation}
This effective metric, unlike $g_{\mu\nu}+h^\R_{\mu\nu}[z]$, does \emph{not} satisfy the vacuum Einstein equation through second order. It does not even satisfy the vacuum equation in the sense that $g_{\mu\nu}+h^\R_{\mu\nu}[\hat z]$ does (i.e., up to dissipation-driven changes in $z^\mu$). One can infer this from the fact that $h^{\R1}_{\mu\nu}$, not $\hat h^{\R1}_{\mu\nu}$, is used in the source for Eq.~\eqref{h2_SC}, meaning $\hat h^{\R2}_{\mu\nu}$ will satisfy $E_{\mu\nu}[\hat h^{\R2}]=2\delta^2 R_{\mu\nu}[h^{\R1},h^{\R1}]$ rather than $E_{\mu\nu}[\hat h^{\R2}]=2\delta^2 R_{\mu\nu}[\hat h^{\R1},\hat h^{\R1}]$. 

Nevertheless, $\hat h^\R_{\mu\nu}$ meets our needs: it is a time-symmetric piece of the retarded field $h_{\mu\nu}[\hat z]$, and $\hat z^\mu$ can be made a geodesic of the associated metric $\btilde g_{\mu\nu}$. I will now verify the latter fact by writing the geodesic equation in the form~\eqref{SC_motion_prev}, but with $\hat z^\mu$ and $\hat h^\R_{\mu\nu}$ in place of $z^\mu$ and $h^\R_{\mu\nu}$, and checking that a circular orbit is a consistent solution. For concreteness, I rewrite the equation here as
\begin{equation}
\frac{D^2 \hat z^\mu}{d\tau^2} = \hat F^\mu[\hat z],\label{SC_motion_con1}
\end{equation}
where $\hat F^\mu[\hat z]$ is given by Eq.~\eqref{force} with the replacement $h^\R_{\mu\nu}\to\hat h^\R_{\mu\nu}$. Explicitly evaluating the covariant derivatives on the left-hand side leads to the algebraic equation
\begin{equation}
\delta^\mu_r\Gamma^r_{uu} = \hat F^\mu[\hat z],\label{SC_motion_con}
\end{equation}
where $\Gamma^\alpha_{uu}\equiv \Gamma^\alpha_{\mu\nu}(\hat z)\hat u^\mu \hat u^\nu$, and I have used the fact that $\Gamma^\mu_{uu}=\delta^\mu_r\Gamma^r_{uu}$. 

To evaluate $\hat F^\mu$, I examine $\hat h^\R_{\mu\nu}$ and its first derivatives on $\hat z^\mu$. Using the facts that $\sum_{m}h_{ni\ell lm}e^{-im\Omega t}Y^{i\ell m}_{\mu\nu}$ must be real and that $Y^{i\ell m*}_{\mu\nu}=(-1)^mY^{i\ell-m}_{\mu\nu}$, we have $h^{*}_{ni\ell m}=(-1)^mh_{ni\ell -m}$. A short calculation then shows that in terms of real quantities,
\begin{multline}
\hat h^{\R n}_{\mu\nu}=\sum_{i\ell}\Bigg[\sum_{m>0}2{\rm Re}\left(H^\R_{ni\ell m}\right)\cos[m(\phi-\Omega t)]P^{i\ell m}_{\mu\nu}\\
+H^R_{ni\ell 0}P^{i\ell0}_{\mu\nu}\Bigg].\label{hsym_decomposition}
\end{multline}
By comparing this with the expansion of $h^{\R n}_{\mu\nu}[\hat z]$, one can easily verify that on $\hat z^\mu$, where $\phi=\Omega t$, the symmetrized field is identical to the nonsymmetrized one:
\begin{equation}
\hat h^{\R n}_{\mu\nu}\bigr|_{\hat z} = h^{\R n}_{\mu\nu}\bigr|_{\hat z}=constant.
\end{equation}
Furthermore, its radial derivative is also equal to that of $h^{\R n}_{\mu\nu}[\hat z]$:
\begin{equation}
\hat h^{\R n}_{\mu\nu,r}\bigr|_{\hat z} = h^{\R n}_{\mu\nu,r}\bigr|_{\hat z}=constant. \label{huu,r}
\end{equation}
However, unlike $h^{\R n}_{\mu\nu}$, it has vanishing $t$ and $\phi$ derivatives on $\hat z^\mu$:
\begin{equation}
\hat h^{\R n}_{\mu\nu,t}\bigr|_{\hat z} = \hat h^{\R n}_{\mu\nu,\phi}\bigr|_{\hat z} = 0.\label{tphi-derivs}
\end{equation}
Also, in a gauge (such as the Lorenz gauge) that respects the system's up-down symmetry we must have that 
\begin{subequations}\label{theta_sym}%
\begin{align}
\hat h^{\R n}_{\mu\nu,\theta}\bigr|_{\hat z}&=h^{\R n}_{\mu\nu,\theta}\bigr|_{\hat z}=0,\\
\hat h^{\R n}_{\mu\theta}\bigr|_{\hat z}&=h^{\R n}_{\mu\theta}\bigr|_{\hat z}=0 \quad\text{for }\mu\neq\theta,
\end{align}
\end{subequations}
since the fields must be invariant under reflection across the equatorial plane.

Now consider the force $\hat F^\mu$. For the sake of comparison, and to see precisely which parts of the force are excluded by using the time-symmetrized field, I will first construct $F^\mu[\hat z]$ from $h^\R_{\mu\nu}$ and only in the final stage make the replacement $h^\R_{\mu\nu}\to\hat h^\R_{\mu\nu}$. After referring to Eq.~\eqref{force} for $F^\mu$ and Eq.~\eqref{u_prev} for $\hat u^\mu$, and utilizing the fact that $\hat u^\rho h^{\R n}_{\mu\nu,\rho} = 0$, I explicitly write the force $F^\mu[\hat z^\mu]$ in Schwarzschild coordinates as
\begin{equation}
F^\mu = -\frac{1}{2}\hat P^{\mu\nu}C_\nu+\frac{1}{2}\hat P^{\mu\nu}h^{\R}_{\nu\rho}g^{\rho\sigma}C_\sigma+O(\e^3),
\end{equation}
where $C_\nu = -h^{\R}_{uu,\nu}-2\Gamma^r_{uu}h^{\R}_{r\nu}$ and $\hat P^{\mu\nu}\equiv g^{\mu\nu}(\hat z)+\hat u^\mu \hat u^\nu$. This force has components
\begin{align}
F^t &= -\frac{1}{2}\hat f^{-1}h^{\R}_{uu,t}+\frac{1}{2}\hat P^{t\beta}\left[2\Gamma^r_{uu}h^{\R}_{r\beta}\right.\nonumber\\
		&\qquad\quad \left.-g^{\gamma\delta}h^{\R}_{\beta\gamma}\left(h^\R_{uu,\delta}
					+2\Gamma^r_{uu}h^{\R}_{r\delta}\right)\right]+O(\e^3),\label{FtV1}\\
F^r &= \frac{1}{2}\hat f\Big[h^{\R}_{uu,r}+2\Gamma^r_{uu}h^{\R}_{rr}\nonumber\\
			&\qquad\quad
			-g^{\alpha\beta}h^{\R}_{r\alpha}\left(h^{\R}_{uu,\beta}+2\Gamma^r_{uu}h^{\R}_{r\beta}\right)\Big]+O(\e^3),\label{FrV1}
\end{align}
where $\hat f\equiv1-2M/\hat r$ and $h^{\R}_{uu}\equiv h^{\R}_{\mu\nu}[\hat z]\hat u^\mu \hat u^\nu$. The $\phi$ component of $F^\mu$ can be found from the orthogonality relation $F^\mu \hat u_\mu=0$, which implies $F^\phi = -\frac{\hat u_t}{\hat u_\phi}F^t$. One can check that $F^\theta$ vanishes by virtue of Eq.~\eqref{theta_sym}.

These expressions can be simplified by appealing to the equation of motion~\eqref{SC_motion_con} (with $F^\mu$ in place of $\hat F^\mu$). We see that $\Gamma^r_{uu}=F^r=\frac{1}{2}\hat f h^{\R1}_{uu,r}+O(\e^2)$. Making that substitution in Eqs.~\eqref{FtV1} and \eqref{FrV1} leads to 
\begin{align}
F^t &= -\frac{1}{2}\e\hat f^{-1}h^{\R1}_{uu,t} -\frac{1}{2}\e^2\Big[\hat f^{-1}h^{\R2}_{uu,t} 
				-\hat P^{t\beta}\big(\hat f^{-1}h^{\R1}_{t\beta}h^{\R1}_{uu,t}\nonumber\\
		&\qquad\qquad\qquad\qquad\quad -\hat r^{-2}h^{\R1}_{\phi\beta}h^{\R1}_{uu,\phi}\big)\Big]+O(\e^3),\label{FtV2}\\
F^r &= \frac{1}{2}\hat f\Big[\e h^{\R1}_{uu,r}+\e^2 \big(h^{\R2}_{uu,r}+\hat f^{-1} h^{\R1}_{tr}h^{\R1}_{uu,t}\nonumber\\
		&\qquad\qquad\qquad\qquad\quad -\hat r^{-2}h^{\R1}_{r\phi}h^{\R1}_{uu,\phi}\big)\Big] +O(\e^3).\label{FrV2}
\end{align}
Clearly, there is no solution to Eq.~\eqref{SC_motion_con} with this force; the left-hand side contains only an $r$ component, while the right-hand side contains $t$ and $\phi$ components. There cannot be a circular orbit accelerated by (the regular part of) the retarded field. 

I now make the change to the time-symmetrized regular field. Imposing Eq.~\eqref{tphi-derivs} in Eqs.~\eqref{FtV2} and \eqref{FrV2}, we find  
\begin{align}
\hat F^t &= \hat F^\phi = \hat F^\theta = O(\e^3),\\
\hat F^r &= \frac{1}{2}\hat f\left(\e \hat h^{\R1}_{uu,r}+\e^2 \hat h^{\R2}_{uu,r}\right)+O(\e^3).\label{hatFr}
\end{align}
With this force, the equation of motion~\eqref{SC_motion_con} clearly does have a solution for $\hat z^\mu$, meaning that $\hat z^\mu$ is a geodesic of the effective metric $\btilde g_{\mu\nu}=g_{\mu\nu}+\hat h^\R_{\mu\nu}$, as desired. I will explore the solution momentarily, but first I comment on how my construction differs from simply neglecting dissipative terms in the equation of motion. On a circular orbit, the components of the force that dissipate energy and momentum are $F^t$ and $F^\phi$; keeping in mind that each component of the force is constant along $\hat z^\mu$, one can easily see that $F^t$ and $F^\phi$ are the only components that change sign under a reversal of the direction of time along the orbit (i.e., $t\to-t$, $\phi\to-\phi$). The radial force $F^r$ is conservative. So in this sense, turning off dissipation consists of setting $F^t=F^\phi=0$ and keeping $F^r$, which allows a precisely circular orbit to be a solution to the equation of motion. Comparing Eq.~\eqref{hatFr} to Eq.~\eqref{FrV2}, we see how this procedures differs from the one I have followed: turning off the dissipative forces leaves terms like $h^{\R1}_{tr}h^{\R1}_{uu,t}$ in the radial force, time-symmetric terms made up of products of time-antisymmetric ones; adopting a geodesic in a time-symmetrized metric, on the other hand, removes those terms. Noting Eq.~\eqref{huu,r}, we see that this is the only difference between the two procedures. In Sec.~\ref{variants} I discuss the result for $\btilde U$ that refers to the conservative dynamics obtained by simply turning off $F^t$ and $F^\phi$. For now, I proceed with the geodesic in $\hat h^\R_{\mu\nu}$.

\subsection{The redshift variable}\label{U_SC}
With the conservative subsystem $(\hat z^\mu, \hat h^\R_{\mu\nu})$ established, I am in a position to fill in the details of Sec.~\ref{preview} to obtain the formula~\eqref{Utilde_SC_prev} for $\btilde U\equiv\frac{dt}{d\btilde\tau}$. 

In my present definition of the conservative dynamics, the four-velocity $\btilde u^\mu=\frac{d\hat z^\mu}{d\btilde\tau}$ is normalized in the time-symmetrized effective metric, and the normalization condition reads $\btilde g_{\mu\nu}[\hat z]\btilde u^\mu \btilde u^\nu = -1$. Using $\btilde u^\mu=\frac{d\tau}{d\btilde\tau}\hat u^\mu$ and $g_{\mu\nu}\hat u^\mu \hat u^\nu=-1$, one finds the ratio between intervals of $\tau$ and $\btilde\tau$ on $\hat z^\mu$ to be
\begin{equation}
\frac{d\tau}{d\btilde\tau} = 1+\frac{1}{2}\hat h^\R_{uu}+\frac{3}{8}\left(\hat h^\R_{uu}\right)^2+O(\epsilon^3),\label{tau_ratio}
\end{equation}
where I have utilized the fact that $\hat h^\R_{\mu\nu}\sim\epsilon$.

Next, from the normalization condition $g_{\mu\nu}\hat u^\mu \hat u^\nu=-1$ and Eq.~\eqref{u_prev}, one finds
\begin{equation}
\hat U^{-2} = \hat f - \hat r^2\Omega^2.\label{U}
\end{equation}

Last, solving the equation of motion~\eqref{SC_motion_con} for the orbital frequency yields 
\begin{multline}
\Omega = \sqrt{\frac{M}{\hat r^3}}\left[1-\frac{\hat F_r \hat r}{2M}(\hat r-3M)\right.\\
\left.-\frac{(\hat F_r)^2 \hat r^2}{8M^2}(\hat r+M)(\hat r-3M)+O(\epsilon^3)\right],\label{Omega_SC}
\end{multline}
where I have used $F^r\sim\epsilon$.

Combining Eqs. \eqref{tau_ratio}, \eqref{U}, and \eqref{Omega_SC}, I obtain a formula for $\btilde U$:
\begin{multline}
\btilde U = (1-3M/\hat r)^{-1/2}\left\{1+\frac{1}{2}(\hat h^\R_{uu}-\hat F_r \hat r)\right.\\
\left.+\frac{1}{8}\left[3(\hat h^\R_{uu})^2-2\hat r\hat F_r\hat h^\R_{uu}-\hat r^2(\hat F_r)^2\right]+O(\epsilon^3)\right\}.\label{Utilde_SC}
\end{multline}
This is the redshift variable in the self-consistent picture and in the definition of conservative dynamics in which the orbit $\hat z^\mu$ is geodesic in the time-symmetrized metric $\btilde g_{\mu\nu}=g_{\mu\nu}+\hat h^\R_{\mu\nu}[\hat z]$. I note that since $\hat h^\R_{\mu\nu}=h^\R_{\mu\nu}$ on $\hat z^\mu$, the hats may be dropped from the regular field in the above formula, recovering Eq.~\eqref{Utilde_SC_prev}. Furthermore, since $\hat h^\R_{\mu\nu,r}=h^\R_{\mu\nu,r}$ on $\hat z^\mu$, we have $\hat F_r = \frac{1}{2}\e h^{\R1}_{uu,r}+O(\e^2)$, so we may remove any explicit reference to the time symmetrization. The only problem with doing so occurs when considering the gauge transformation of $\btilde U$; I postpone that discussion to Sec.~\ref{gauge_GW}.

\section{Gralla-Wald picture: an expanded worldline}\label{GW}
The reason for transitioning to a Gralla-Wald picture should now be clear: If one tried to calculate the value of $\btilde U$ numerically by solving Eqs.~\eqref{h1_SC} and \eqref{h2_SC} with the motion of the puncture determined by Eq.~\eqref{SC_motion_con1}, to avoid numerical error driving the orbit away from circularity, one would have to constrain the orbit \emph{a priori} to be circular. But to do so one would have to know the correct initial conditions for the position and velocity of that orbit. In other words, one would need the relationship between $\hat r$ and $\Omega$ in Eq.~\eqref{circ_prev}, which [from Eq.~\eqref{Omega_SC}] would require knowing the correct radial force in advance. A Gralla-Wald scheme circumvents this challenge.

\subsection{Formalism}\label{GW_formalism}
In the Gralla-Wald picture, the accelerated worldline is expanded in a power series around some reference geodesic of $g_{\mu\nu}$. One could begin with an expansion of the inspiraling worldline and then extract the conservative dynamics, and in fact I present that approach in Appendix~\ref{dissipation}. But here I begin instead with the conservative orbit $\hat z^\mu$. Following Sec.~\ref{preview}, I expand it as
\begin{equation}
\hat z^\mu(t,\epsilon) = z^\mu_0(t)+\epsilon\hat z^\mu_1(t)+\epsilon^2\hat z^\mu_2(t)+O(\epsilon^3),
\end{equation}
where the zeroth-order worldline is 
\begin{equation}
z_0^\mu = \left\lbrace t, r_0, \frac{\pi}{2}, \Omega_0 t\right\rbrace,\label{z0}
\end{equation}
a circular, background geodesic with frequency
\begin{equation}
\Omega_0 = \sqrt{\frac{M}{r_0^3}}\label{Omega0}
\end{equation}
and four-velocity
\begin{equation}
u_0\equiv \frac{dz_0^\mu}{d\tau_0}= U_0\{1,0,0,\Omega_0\}.
\end{equation}
From Eq.~\eqref{U}, we have
\begin{equation}
U^{-2}_0 = f_0 - r_0^2\Omega_0^2=1-\frac{3M}{r_0},\label{U0}
\end{equation}
where $f_0\equiv 1-2M/r_0$.

In Sec.~\ref{preview}, I chose $z^\mu_0$ to be the circular geodesic with frequency $\Omega_0=\Omega$. I will eventually make that same choice here, but for the moment, to keep the discussion general, I leave the frequency $\Omega_0$ arbitrary. The corrections to $z_0^\mu$ are then, generically, $\hat z_{n}^\mu \equiv \{0,\ \hat r_n,\ 0,\ \Omega_n t\}\equiv\frac{1}{n!}\frac{d^n\hat z^\mu}{d\epsilon^n}|_{\epsilon=0}$. To interpret these quantities more formally, note that $\hat z^\mu(t,\epsilon)$ parametrizes a two-dimensional surface that is bounded on one side by $\hat z^\mu(t,0)=z_0^\mu(t)$. $\hat z^\mu_1$ is the directional derivative $\frac{\partial\hat z^\mu}{\partial\epsilon}(t,\epsilon=0)$ along a curve of increasing $\epsilon$ and fixed $t$ in this surface; therefore, it is a vector field on $z^\mu_0$, transforming in the ordinary way as a vector there. $\hat z^\mu_2$, on the other hand, is a second derivative along this curve, rather than a first; therefore, it is simply a collection of four scalar fields on $z^\mu_0$, rather than a vector field. 

When the worldline is expanded in this way, the fields $h^n_{\mu\nu}[\hat z]$  must also be expanded (along with $h^{\R n}_{\mu\nu}[\hat z]$,  $\hat h^n_{\mu\nu}[\hat z]$, etc.). Writing 
\begin{equation}
h_{\mu\nu}=\e h^1_{\mu\nu}[z_0+\e \hat z_1+\ldots]+\e^2h^2_{\mu\nu}[z_0+\e \hat z_1+\ldots]
\end{equation}
and then expanding the functional dependence yields a field of the form
\begin{equation}
h_{\mu\nu}=\e h^{1}_{\mu\nu}[z_0]+\e^2(h^{2}_{\mu\nu}[z_0]+\delta h^1_{\mu\nu}[z_0,\hat z_1])+O(\e^3).
\end{equation}
The term $\delta h^1_{\mu\nu}$ comes from functional differentiation of $h^1_{\mu\nu}$. I incorporate that term into a new second-order field, $h^{2({\rm GW})}_{\mu\nu}$, to arrive at
\begin{equation}
h_{\mu\nu}=\e h^{1}_{\mu\nu}[z_0]+\e^2 h^{2({\rm GW})}_{\mu\nu}[z_0,\hat z_1]+O(\e^3).
\end{equation}
The first term is the same functional as in the self-consistent picture, but now evaluated as a functional of $z_0^\mu$; this approximation is often made in the self-force literature at first order. The second term, $h^{2({\rm GW})}_{\mu\nu}[z_0,\hat z_1]$, is a different functional than in the self-consistent case, due to its inclusion of $\delta h^1_{\mu\nu}$. Its form will be made clear by analyzing its singular and regular pieces.

First consider the singular field. Substituting the expansion~\eqref{expanded_zhat} into the schematic expressions~\eqref{hS1_SC_schematic} and~\eqref{hS2_SC_schematic}, one finds that near the object, the singular field in the Gralla-Wald picture takes the form 
\begin{equation}
h^{\S}_{\mu\nu}=\e h^{\S1}_{\mu\nu}[z_0]+\e^2h^{\S2({\rm GW})}_{\mu\nu}[z_0,\hat z_1]+O(\e^3),
\end{equation}
 with
\begin{align}
h^{\S1}_{\mu\nu}[z_0]&\sim \frac{m}{|x^\alpha-z^\alpha_0|} + O(|x^\alpha-z^\alpha_0|^0),\label{hS1_GW_schematic}\\
h^{\S2({\rm GW})}_{\mu\nu}[z_0,\hat z_1]&\sim \frac{m^2+m\hat z^\mu_{1\perp}}{|x^\alpha-z^\alpha_0|^2} 
		+ \frac{\delta m_{\mu\nu}+mh^{\R1}_{\mu\nu}}{|x^\alpha-z^\alpha_0|}	\nonumber\\
		&\quad + O(\ln|x^\alpha-z^\alpha_0|).\label{hS2_GW_schematic}
\end{align}
Here $|x^\alpha-z_0^\mu|$ represents spatial distance from $z^\mu_0$, and $\hat z_{1\perp}^\alpha\equiv (g^\alpha{}_\beta +u_0^\alpha u_{0\beta})\hat z_1^\beta$ is the piece of $\hat z^\alpha_1$ orthogonal to the zeroth-order worldline. The singular field now diverges on $z_0^\mu$ rather than $\hat z^\mu$; the correction to the motion, instead of shifting the location of the divergence, now appears explicitly as a term in the field. As in the self-consistent case, the local expansions~\eqref{hS1_GW_schematic} and \eqref{hS2_GW_schematic} can be found in explicit, covariant form in Ref.~\cite{Pound-Miller:14}. 

In the $1/|x^\alpha-z_0^\alpha|$ term in $h^{\S2}_{\mu\nu}$, the regular field $h^{\R1}_{\mu\nu}$ is both evaluated at $z_0^\mu$ and a functional of $z_0^\mu$; the functional $h^{\R1}_{\mu\nu}[\hat z]$ is approximated by $h^{\R1}_{\mu\nu}[z_0^\mu]$. The expansion of the worldline also alters the tensor $\delta m_{\mu\nu}$, which now reads
\begin{align}\label{dm_GW}
\delta m_{\alpha\beta} &= \frac{1}{3}m\left(2h^{\R1}_{\alpha\beta}+g_{\alpha\beta}g^{\mu\nu}h^{\R1}_{\mu\nu}\right)\nonumber\\
&\quad +m(g_{\alpha\beta}+2u_{0\alpha} u_{0\beta})u_0^\mu u_0^\nu h^{\R1}_{\mu\nu}\nonumber\\
&\quad +4mu_{0(\alpha}\left(h^{\R1}_{\beta)\mu}u_0^\mu+2\frac{D\hat z^\perp_{1\beta)}}{d\tau_0}\right),
\end{align}
where $h^{\R1}_{\mu\nu}$ is again both evaluated on $z_0^\mu$ and a functional of $z_0^\mu$.

Expanding the singular field in this way leaves intact most of the puncture scheme described in Sec.~\ref{SC_formalism}, but for two important modifications:
\begin{itemize}
\item the punctures $h^{\P n}_{\mu\nu}$ move on $z_0^\mu$, not on $\hat z^\mu$,
\item the second-order puncture includes terms proportional to $\hat z^\mu_1$.
\end{itemize}
The first of these two changes renders a calculation of the conservative dynamics far simpler than in the self-consistent picture. Rather than having to somehow predetermine the relationship between orbital frequency and radius of the perturbed orbit, one can now choose a background geodesic $z_0^\mu$ howsoever one likes, and the puncture moves on that geodesic at both first and second order. The puncture scheme, rather than consisting of a system of coupled equations, becomes a sequence of equations: After specifying the background geodesic, one can calculate the first-order fields $h^1_{\mu\nu}[z_0]$ and $h^{\R1}_{\mu\nu}[z_0]$ by solving the new version of Eq.~\eqref{h1_SC},
\begin{subequations}\label{h1_GW}%
\begin{align}
E_{\mu\nu}[h^{\res1}] &= -E_{\mu\nu}[h^{\P1}]\equiv S_{\mu\nu}^{\rm 1eff} & \text{inside }\Gamma,\\
E_{\mu\nu}[h^{1}] &= 0 & \text{outside }\Gamma,
\end{align}
\end{subequations}
with the puncture now moving on $z^\mu_0$, or by solving the new version of Eq.~\eqref{h1_GW_point},
\begin{align}
E_{\mu\nu}[\bar h^1] &= -16\pi \int_{z_0} mu_{0\mu} u_{0\nu} \frac{\delta^4(x^\alpha-z_0^\alpha)}{\sqrt{-g}}d\tau_0.\label{h1_GW_point}
\end{align}
Next, one can use the linear-in-$\e$ term in the equation of motion~\eqref{SC_motion_prev} [or Eq.~\eqref{SC_motion_con}] to find the correction $\hat z_1^\mu$. After that, one can solve the second-order field equation, 
\begin{subequations}\label{h2_GW}%
\begin{align}
E_{\mu\nu}[h^{\res2}] &= 2\delta^2R_{\mu\nu}[h^1,h^1]- E_{\mu\nu}[h^{\P2}]\hspace{-10pt} &\nonumber\\
											& \equiv S^{2\rm eff}_{\mu\nu}  & \text{inside }\Gamma,\label{h2_GW_in}\\
E_{\mu\nu}[h^2] &= 2\delta^2R_{\mu\nu}[h^1,h^1] & \text{outside }\Gamma,\label{h2_GW_out}
\end{align}
\end{subequations}
with the puncture still moving on $z^\mu_0$. Finally, one can use the quadratic-in-$\e$ term in the equation of motion to find the correction $\hat z_2^\mu$. In the above, I have omitted the label ``(GW)'' on $h^2_{\mu\nu}$ for compactness.

From this puncture scheme, and the form of the puncture, we see that the first-order regular field is now what you would obtain by taking the first-order puncture in the self-consistent picture and setting it moving on $z_0^\mu$. The second-order regular field here is what you would obtain by taking the second-order puncture in the self-consistent picture and setting it moving on $z_0^\mu$, \emph{plus} the regular field generated from the new terms proportional to $\hat z^\mu_1$ in the puncture. 

It is illuminating to consider these expansions from the perspective of the first-order, point-mass stress-energy tensor~\eqref{T1}. Expanding $T^1_{\mu\nu}[\hat z]$ about $z_0^\mu$ leads to two terms involving $\hat z_1^\mu$. First, the expansion of the Dirac $\delta$ function produces a $\delta'$ source, leading to the term $\sim\hat z_1^\mu/|x^\alpha-z^\alpha_0|^2$ in Eq.~\eqref{hS2_GW_schematic}. Second, the expansion of $u^\mu$ about $u_0^\mu$ in $T^1_{\mu\nu}[\hat z]$ produces a $\delta$ source proportional to $\frac{Dz_1^\mu}{d\tau_0}$, leading to the new term in $\delta m_{\mu\nu}/|x^\alpha-z_0^\alpha|$ shown in Eq.~\eqref{dm_GW}. The fact that only $\hat z_{1\perp}^\mu$ contributes to the singular field, rather than the entirety of $\hat z_{1}^\mu$, can be seen from a careful analysis of the change of integration variable from $\tau$ to $\tau_0$ in $T^1_{\mu\nu}$~\cite{Pound:14a}. 

Similarly, to better understand how the regular field is altered by the expansion of $\hat z^\mu$, one can refer to the explicit functional~\eqref{hR1_Greens}. Substituting the expansion into the right-hand side of that equation leads to $h^{\R1}_{\mu\nu}[\hat z]=h^{\R1}_{\mu\nu}[z_0]+\e \delta h^{\R1}_{\mu\nu}[z_0,\hat z_1]+O(\e^2)$, where
\begin{multline}
\delta h^{\R1}_{\mu\nu} = 4m\int_{z_0} \Big(\bar G^\R_{\mu\nu\alpha'\beta';\gamma'}u_0^{\alpha'} u_0^{\beta'}\hat z_{1\perp}^{\gamma'}\\
																		+2\bar G^\R_{\mu\nu\alpha'\beta'}\frac{D\hat z_{1\perp}^{\alpha'}}{d\tau} u_0^{\beta'}\Big) d\tau_0.
\end{multline}
The primed indices refer to the tangent space at $x'^{\mu}=z^\mu_0(\tau_0)$, and again, accounting for the change of integration variable explains the fact that only the perpendicular piece of $\hat z_{1}^{\alpha}$ appears~\cite{Pound:14a}. Given this expansion, we ascertain that the first- and second-order regular fields in the Gralla-Wald picture have the form 
\begin{align}
h^{\R1({\rm GW})}_{\mu\nu} &\equiv h^{\R1}_{\mu\nu}[z_0],\\
h^{\R2({\rm GW})}_{\mu\nu} &\equiv h^{\R2}_{\mu\nu}[z_0]+\delta h^{\R1}_{\mu\nu}[z_0,\hat z_1],
\end{align}
with $\delta h^{\R1}_{\mu\nu}$ as given above.

So that I can say more about the problem at hand, allow me to return to the choice of $z_0^\mu$ made in Sec.~\ref{preview}, where $\Omega=\Omega_0$ and the correction to the position in Eq.~\eqref{expanded_zhat} is purely radial. If that choice is made, the expansion of the worldline can be written as
\begin{equation}\label{zhat-dr}
\hat z^\mu=z^\mu_0+\epsilon\hat r^\mu_1+\epsilon^2\hat r^\mu_2+O(\epsilon^3),
\end{equation}
where $\hat r_{n}^\mu\equiv\delta^\mu_r\hat r_n$. The zeroth-order four-velocity is then proportional to the same Killing vector as is $u^\alpha$,
\begin{equation}
u^\alpha_0 = U_0 k^\alpha,
\end{equation}
and the perturbations retain their helical symmetry,
\begin{equation}
\mathcal{L}_k h^{\R n({\rm GW})}_{\mu\nu} = 0,\label{hRC_symmetry}
\end{equation}
and the same for the retarded and singular fields. This can be gleaned from Eq.~\eqref{h1_helical}, for example, by observing that only the dependence on $\hat r$ is expanded, leaving the $t$ and $\phi$ dependence unaltered. On the zeroth-order worldline the helical symmetry reduces to
\begin{equation}
u_0^\rho  h^{\R n({\rm GW})}_{\mu\nu,\rho} = 0.\label{hRC_const}
\end{equation}

In deriving the expansion of $\btilde U$ below, I will be interested in the regular field (and its derivatives) evaluated on the worldline---for example, as it appears in the equation of motion~\eqref{force}. This means I will require an expansion of the regular field on $\hat z^\mu$ in the self-consistent picture about the regular field on $z_0^\mu$ in the Gralla-Wald picture. To make that expansion more transparent, I switch notation from $h^{\R n}_{\mu\nu}[\hat z]$ to $h^{\R n}_{\mu\nu}(x;\hat z)$, where $x^\mu$ is the point at which the field is evaluated. Expanding $h^{\R n}_{\mu\nu}(\hat z;\hat z)$ around $(z_0;z_0)$, using Eq.~\eqref{zhat-dr}, leads to 
\begin{multline}
h^{\R n}_{\mu\nu}(\hat z;\hat z) = h^{\R n}_{\mu\nu}(z_0;z_0)+\epsilon \Big[\hat r_1h^{\R n}_{\mu\nu,r}(z_0;z_0)\\
							+\delta h^{\R n}_{\mu\nu}(z_0;z_0,\hat r_1)\Big]+O(\epsilon^2).\label{expanded_hRn}
\end{multline}
The first term in the square brackets accounts for the shift in the field point from $z^\mu_0(t)$ to $\hat z^\mu(t)$, while the second term accounts for the shift in the source orbit from $z^\mu_0$ to $\hat z^\mu$. Equation~\eqref{expanded_hRn} implies that the total regular field on the accelerated worldline can be expanded as  
\begin{multline}
h^{\R}_{\mu\nu}(\hat z;\hat z) = \e h^{\R1({\rm GW})}_{\mu\nu}(z_0)+\e^2 \Big[h^{\R1 ({\rm GW})}_{\mu\nu,r}(z_0)\hat r_1\\
							 										+h^{\R2({\rm GW})}_{\mu\nu}(z_0)\Big]+O(\epsilon^2),\label{expanded_hR}
\end{multline}
where I have suppressed the functional dependences on the right-hand side. Finally, substituting these results into the force~\eqref{hatFr} leads to an expansion $\hat F^r = \e F_1^r+\e^2 F^r_2$, where
\begin{align}
\hat F_1^r &= \frac{1}{2}f_0 \hat h^{\R1({\rm GW})}_{u_0u_0,r},\label{F1}\\
\hat F_2^r &= \frac{1}{2}f_0 \hat h^{\R2({\rm GW})}_{u_0u_0,r}+\frac{M\hat r_1}{r_0^2} \hat h^{\R1({\rm GW})}_{u_0u_0,r}+\frac{1}{2}f_0 \hat h^{\R1({\rm GW})}_{u_0u_0,rr}\hat r_1.\label{F2}
\end{align}
Here I have introduced $\hat h^{\R n({\rm GW})}_{u_0u_0}\equiv h^{\R n({\rm GW})}_{\mu\nu}u^\mu_0u^\nu_0$ and $h^{\R n({\rm GW})}_{u_0u_0,\rho}\equiv h^{\R n({\rm GW})}_{\mu\nu,\rho}u^\mu_0u^\nu_0$.

For the sake of notational simplicity, from this point forward I will omit the `(GW)' label, and $h^{\R n}_{\mu\nu}$ will always represent $h^{\R n{\rm (GW)}}_{\mu\nu}$.

\subsection{Corrections to the orbital radius}\label{expanded_motion}
In order to obtain my final expansion of $\btilde U$, I must first solve the equations of motion for the corrections $\hat z^\mu_n$ to the motion; otherwise I will be left with an unhelpful expression in terms of $\hat r_1$, for example. In this section I do just that, finding $\hat z^\mu_n$ by substituting the expansion~\eqref{expanded_zhat} into the equation of motion~\eqref{SC_motion_con}. To illustrate the freedom in the Gralla-Wald picture, I momentarily delay the choice $\Omega=\Omega_0$. 

For convenience, I restate the equation of motion here: 
\begin{equation}
\Gamma^r_{\mu\nu}(\hat r)u^\mu u^\nu = \e \hat F_1^r+\e^2 \hat F_2^r+O(\e^3).\label{EqMot}
\end{equation}
$\hat F_1^r$ and $\hat F_2^r$ are given in Eqs.~\eqref{F1} and~\eqref{F2}, but those concrete expressions will not be needed for the present analysis.

The zeroth-order term in Eq.~\eqref{EqMot} reads $\Gamma^\alpha_{\mu\nu}(r_0)k^\mu k^\nu = 0$, the solution of which is the familiar formula~\eqref{Omega0}. 

The first-order term in Eq.~\eqref{EqMot} reads
\begin{equation}\label{EqMot1}
\Gamma^r_{\mu\nu,r}(r_0)u_0^\mu u_0^\nu \hat r_1 +2\Omega_0\Omega_1U_0^2 \Gamma^r_{\phi\phi}(r_0) = \hat F_1^r.
\end{equation}
Even if $\hat F_1^r \equiv 0$, this equation has a nontrivial solution relating $\hat r_1$ to $\Omega_1$. That solution corresponds to a small shift to another circular geodesic of slightly different radius, unrelated to the self-force or any physical perturbation. I eliminate it by setting $\Omega_1=0$. This same freedom resides at every order, and I eliminate it by making the promised choice 
\begin{equation}
\Omega = \Omega_0.
\end{equation}

Rather than choosing $\Omega = \Omega_0$, one could use the freedom in the expansion to make the alternative choice $\hat r(\epsilon)=r_0$, or some choice of relation $\Omega=\Omega(\hat r)$. In those cases, one would have nonzero shifts $\Omega_n$ in the orbital frequency, such that $\Omega\neq \Omega_0$. The different choices correspond to different families of orbits---and to different families of spacetimes (parametrized by $\e$). In the case that $\Omega = \Omega_0$, each member of the family contains a compact object orbiting at a physical frequency $\Omega$. In the case that $\Omega \neq \Omega_0$, different members of the family have physically different frequencies. There are two reasons for choosing $\Omega = \Omega_0$ in the present analysis: it means that the coordinate-dependent (though gauge-independent, as discussed in Sec.~\ref{gauge_GW}) radius $r_0$ can be written in terms of the physical frequency as $r_0=(M\Omega^{-2})^{1/3}$; it also means that the corrections $\hat z_{n>0}^\mu$ are purely radial and constant in time. With a generic choice of relation $\Omega=\Omega(\hat r)$, $r_0$ would be nontrivially related to $(M\Omega^{-2})^{1/3}$. More problematically, the first-order correction to the motion, $\hat z_1^\mu$, would include $\phi_1(t)=\Omega_1 t$. Terms growing linearly in time would then appear in the metric, corresponding to expanding $\Omega(\e)$ in, e.g., Eq.~\eqref{h1_helical}, and the equation of motion would become time dependent, substantially complicating the analysis. However, for some purposes,  one would \emph{require} a family of spacetimes of differing frequency. For example, if one wished to define the self-force-induced shift in frequency of the ISCO, one would consider a family in which the object is at the ISCO at each $\e$. In such cases, one would have to use a slightly different formalism to bypass the inconvenient growth in time.

I now return to Eq.~\eqref{EqMot1} from my digression. Using Eqs.~\eqref{Omega0} and \eqref{U0}, I find the first-order shift in the orbital radius due to the self-force to be
\begin{equation}\label{r1}
\hat r_1 = -\frac{r_0^3}{3M}\frac{(r_0-3M)}{(r_0-2M)}\hat F_1^r.
\end{equation}
It follows from this and Eq.~\eqref{U} that
\begin{equation}
\hat U=U_0+O(\epsilon^2),
\end{equation}
and so
\begin{equation}
\hat u^\mu=u^\mu_0+O(\epsilon^2).\label{u_expansion}
\end{equation}

The second-order shift is similarly found from the second-order term in Eq.~\eqref{EqMot}. That equation reads
\begin{equation}\label{EqMot2}
\Gamma^r_{\mu\nu,r}(r_0)u_0^\mu u_0^\nu \hat r_2 +\frac{1}{2}U_0^2\Gamma^r_{tt,rr}(r_0)\hat r_1^2 = \hat F_2^r,
\end{equation}
and its solution is
\begin{equation}\label{r2}
\hat r_2 = -\frac{r_0^3}{3M}\frac{(r_0-3M)}{(r_0-2M)}\hat F^r_2 +\frac{\hat r_1^2}{r_0}\frac{r_0-4M}{r_0-2M}.
\end{equation} 
One can check the consistency of these results for $\hat r_1$ and $\hat r_2$ by substituting them into Eq.~\eqref{Omega_SC}, which returns $\Omega=\Omega_0$ through second order, as required.

\subsection{The redshift variable}\label{Utilde_GW}
I now turn to the expansion of $\btilde U$. Substituting the expansions~\eqref{zhat-dr}, \eqref{expanded_hR}, \eqref{u_expansion}, and 
\begin{align}
\hat F_{r}=\hat f^{-1}\hat F^r &=\epsilon f^{-1}_0 \hat F_1^r+\epsilon^2 f^{-1}_0\left(\hat F_2^r-\frac{2M}{f_0r_0^2} \hat r_1\hat F_1^r\right)\nonumber\\
	&\quad+O(\epsilon^3) 
\end{align}
into Eq.~\eqref{Utilde_SC}, one finds
\begin{align}
\btilde U &= U_0\left\{1+\frac{1}{2}\epsilon \hat h^{\R1}_{u_0u_0}+\epsilon^2\left[\frac{1}{2}(\hat h^{\R2}_{u_0u_0}+\hat h^{\R1}_{u_0u_0,r}\hat r_1)\right.\right.\nonumber\\
&\quad\left.\left.+\frac{3}{8}(\hat h^{\R1}_{u_0u_0})^2+\frac{r_0^2}{6M}(r_0-3M)(\hat F_{1r})^2\right]+O(\epsilon^3)\right\}.\label{Utilde_GWv1}
\end{align}

To simplify this expression, I eliminate $\hat h^{R1}_{u_0u_0,r}$ and $\hat r_1$ by making use of Eqs.~\eqref{F1} and \eqref{r1}, leading to
\begin{multline}
\btilde U = U_0\left\{1+\frac{1}{2}\epsilon \hat h^{R1}_{u_0u_0}+\epsilon^2\left[\frac{1}{2}\hat h^{R2}_{u_0u_0}+\frac{3}{8}(\hat h^{R1}_{u_0u_0})^2\right.\right.\\
\left.\left.-\frac{r_0^2}{6M}(r_0-3M)(\hat F_{1r})^2\right]+O(\epsilon^3)\right\}\label{Utilde_GWv2},
\end{multline}
where, recall, $\hat F_{1r} = \frac{1}{2}\hat h^{\R1}_{u_0u_0,r}$. 

As noted in Sec.~\ref{U_SC}, $\hat h^{\R n}_{\mu\nu}$ can be replaced with $h^{\R n}_{\mu\nu}$ in Eq.~\eqref{Utilde_GWv2}, including within $\hat F_{1r}$, thereby recovering the previewed equation~\eqref{Utilde_GW_prev}. With the present construction of the conservative dynamics, that formula can be taken to describe the ratio $dt/d\btilde\tau$ along the circular orbit $\hat z^\mu$ that is a geodesic of the time-symmetrized effective metric $\btilde g_{\mu\nu}=g_{\mu\nu}+\hat h^{\R}_{\mu\nu}$.

\section{Gauge transformations}\label{gauge_GW}
Only one step remains: to show that $\btilde U$ is gauge invariant. Thus far I have restricted my attention to the Lorenz gauge. I now describe the effects of a transformation to another gauge. Before specializing to the conservative system comprising $\hat z^\mu$ and $\hat h_{\mu\nu}$, I give a general description of transformations at second order. I use standard results from, e.g., Ref.~\cite{Bruni-etal:96}.

\subsection{Transformation laws}
In the self-consistent formalism, the starting point is the transformation of the worldline itself. Under a smooth gauge transformation generated by $\epsilon\xi_1^\mu$ and $\epsilon^2\xi_2^\mu$, the coordinates $z^\mu$ on the worldline transform according to
\begin{align}
z^\mu\to z'^{\mu} &= z^\mu-\epsilon \xi^\mu_1(z)-\epsilon^2 \left[\xi^\mu_2(z)-\frac{1}{2}\xi^\nu_1(z)\partial_\nu\xi_1^\mu(z)\right]\nonumber\\
 &\quad +O(\epsilon^3).\label{z_transformation}
\end{align}
When the worldline is expanded in a Taylor series, the terms in its expansion transform according to
\begin{align}
z_0^\mu &\to z_0^\mu, \\
z_1^\mu &\to  z_1'^{\mu} =  z_1^\mu - \xi_1^\mu(z_0), \label{z1_transformation}\\
z_2^\mu &\to  z_2'^{\mu} =  z_2^\mu - \xi_2^\mu(z_0)+\frac{1}{2}\xi^\nu_1(z_0)\partial_\nu\xi_1^\mu(z_0)\nonumber\\
		&\qquad\qquad-z_1^\nu\partial_\nu\xi^\mu_1(z_0),\label{z2_transformation}
\end{align}
The laws for $z^\mu_n$ follow from Eq.~\eqref{z_transformation} by expanding both sides of the equality about the zeroth-order worldline $z_0^\mu$ and equating coefficients of powers of $\epsilon$. They can also be found from a more detailed differential-geometric analysis. 

We can see that a gauge transformation acts quite differently in the two pictures. In the self-consistent picture, a gauge transformation shifts the curve on which the singular field diverges. In the Gralla-Wald picture, on the other hand, the curve $z_0^\mu$ on which the singular field diverges is trivially invariant in the usual sense of any zeroth-order quantity in perturbation theory. Instead, the gauge transformation alters the fields $ z_{1}^\mu$, $ z_2^\mu$, $\ldots$, that live on $z_0^\mu$.

In either picture, under the same smooth gauge transformation, $h^n_{\mu\nu}$ transforms as $h^n_{\mu\nu}\to h'^n_{\mu\nu}=h^n_{\mu\nu}+\Delta h^n_{\mu\nu}$, where\footnote{Reference~\cite{Pound:14b} discusses the subtleties that arise when applying this formula in a self-consistent scheme.}
\begin{subequations}\label{Dh}%
\begin{align}
\Delta h^1_{\mu\nu} &=\mathcal{L}_{\xi_1} g_{\mu\nu},\\
\Delta h^2_{\mu\nu} &=\mathcal{L}_{\xi_2} g_{\mu\nu}+\mathcal{L}_{\xi_1} h^{1}_{\mu\nu} + \frac{1}{2}\mathcal{L}^2_{\xi_1} g_{\mu\nu}.
\end{align}
\end{subequations}
Obviously this transformation is valid only off the worldline, where the fields are smooth. Now we must apportion $\Delta h^n_{\mu\nu}$ between the singular and regular fields, which takes some thought. My guiding principle is this: I wish to define the regular field $h'^{\R}_{\mu\nu}$ in the new gauge such that the equation of motion~\eqref{SC_motion_prev} remains valid, with $z^\mu$ replaced by $z'^\mu$ and $h^\R_{\mu\nu}$ by $h'^\R_{\mu\nu}$. In its geodesic form~\eqref{geodesic_form_prev}, the equation of motion is manifestly invariant under a general smooth coordinate transformation. Therefore, when $\tilde g_{\mu\nu}$ is split into the background $g_{\mu\nu}$ and the perturbation $h^{\R}_{\mu\nu}$, the equation in the form~\eqref{SC_motion_prev} must be invariant when $z^\mu$ transforms as \eqref{z_transformation} and $h^\R_{\mu\nu}$ transforms as any smooth perturbation would. Accordingly, I define the regular field in the new gauge as $h'^{\R n}_{\mu\nu}=h^{\R n}_{\mu\nu}+\Delta h^{\R n}_{\mu\nu}$, where
\begin{subequations}\label{DhR}%
\begin{align}
\Delta h^{\R1}_{\mu\nu} &=\mathcal{L}_{\xi_1} g_{\mu\nu},\\
\Delta h^{\R2}_{\mu\nu} &=\mathcal{L}_{\xi_2} g_{\mu\nu}+\mathcal{L}_{\xi_1} h^{\R1}_{\mu\nu} 
		+ \frac{1}{2}\mathcal{L}^2_{\xi_1} g_{\mu\nu}.
\end{align}
\end{subequations}
With these definitions, the effective metric $\btilde g_{\mu\nu}=g_{\mu\nu}+h^\R_{\mu\nu}$ transforms just as any other smooth metric, and it retains its properties in the new gauge: it is a $C^\infty$ solution to the vacuum Einstein equation, and the orbit is geodesic in it. The transformation law for $h^{\R n}_{\mu\nu}$ leaves the singular field to transform as $h^{\S n}_{\mu\nu}\to h'^{\S n}_{\mu\nu}=h^{\S n}_{\mu\nu}+\Delta h^{\S n}_{\mu\nu}$, where
\begin{subequations}\label{DhS}%
\begin{align}
\Delta h^{\S1}_{\mu\nu} &=0,\\
\Delta h^{\S2}_{\mu\nu} &=\mathcal{L}_{\xi_1} h^{\S1}_{\mu\nu}.\label{DhS2}
\end{align}
\end{subequations}
Since only certain combinations of $h^{\R}_{\mu\nu}$ and its derivatives appear in the equation of motion, it might be possible to put more of $\Delta h^n_{\mu\nu}$ into $\Delta h^{\S n}_{\mu\nu}$ without spoiling the invariance of Eq.~\eqref{SC_motion_prev}. However, the above definitions are the most natural.

Since the equation of motion is invariant, one can see that much of the analysis in the preceding sections remains valid in any gauge that is smoothly related to Lorenz. In fact, the entirety of the analysis remains valid so long as we restrict ourselves to a class of gauges that preserve the system's helical symmetry, by which I mean that the orbit retains a manifestly circular, equatorial form, as in Eq.~\eqref{circ_prev}, and that the helical Killing vector retains the form~\eqref{helical_prev}. These conditions can be guaranteed by restricting $\xi^\mu_n$ to satisfy 
\begin{equation}
0 = \mathcal{L}_{\xi_n} k^\mu = -k^\nu \partial_\nu \xi_n^\mu,\label{xi_symmetry}
\end{equation}
and $\xi_n^\theta=0$. On the worldline, Eq.~\eqref{xi_symmetry} reduces to $\frac{d\xi^\mu_n}{d\tau}=0$ (or $\frac{d\xi^\mu_n}{d\tau_0}=0$, in the Gralla-Wald picture); the gauge vector must be constant on the worldline. The condition $\xi_n^\theta=0$ keeps the worldline in the equatorial plane of the background and preserves the metric perturbation's symmetry about that plane. 

Within this class of gauges, $\hat z^\mu$ transforms according to Eq.~\eqref{z_transformation}; the transformation of $\hat z^\mu$ corresponds to a constant shift in the orbit's radius and a shift of its initial azimuthal angle. A natural transformation law for $\hat h^\R_{\mu\nu}$ can be found by again demanding that the equation of motion takes the same form in the new gauge. $\hat z^\mu$ satisfies the geodesic equation in the effective metric $\btilde g_{\mu\nu}=g_{\mu\nu}+\hat h^\R_{\mu\nu}$, so by the same argument as above, I let $\hat h^{\R}_{\mu\nu}$ transform according to Eq.~\eqref{DhR}, with $\hat h^{\R n}_{\mu\nu}$ replacing $h^{\R n}_{\mu\nu}$ on both the left- and right-hand sides. 

\subsection{Invariance of the redshift variable}
All the rules of transformation are now established, and so I turn to the actual quantities appearing in Eq.~\eqref{Utilde_GWv2}. First, we have
\begin{equation}\label{DhR1}
\Delta \hat h^{\R1}_{u_0u_0} = \Delta h^{\R1}_{u_0u_0} = 2\frac{d}{d\tau_0}(\xi_{1\mu}u_0^\mu) =0,
\end{equation}
where I have now restricted $\xi_n^\mu$ to satisfy Eq.~\eqref{xi_symmetry}. Next, 
\begin{equation}
\Delta \hat h^{\R2}_{u_0u_0} = \hat h^{\R1}_{u_0u_0,\rho}\xi^\rho_1+\frac{3M}{r_0^2}\frac{(\xi_1^r)^2}{r_0-3M},
\end{equation}
where I have again used Eq.~\eqref{xi_symmetry}. I simplify the result for $\Delta \hat h^{\R2}_{u_0u_0}$ by noting that on $z_0^\mu$ we have $\hat h^{\R1}_{u_0u_0,r}=2\hat F_{1r}$ and $\hat h^{\R1}_{\mu\nu,t}= 0= \hat h^{\R1}_{\mu\nu,\phi}$, which leads to
\begin{equation}
\Delta \hat h^{\R2}_{u_0u_0}= 2\hat F_{1r}\xi^r_1+\frac{3M}{r_0^2}\frac{(\xi_1^r)^2}{r_0-3M}.\label{DhR2}
\end{equation}
This result does \emph{not} hold for the nonsymmetrized $\Delta h^{\R2}_{u_0u_0}$, because $h^{\R1}_{u_0u_0,t}\neq 0\neq h^{\R1}_{u_0u_0,\phi}$.

Using $\hat F_{1r}=\frac{1}{2}\hat h^{\R1}_{u_0u_0,r}$ and the transformation law for $\hat h^{\R1}_{\mu\nu}$, and again appealing to Eq.~\eqref{xi_symmetry}, I next find that $\hat F_{1r}$ transforms as
\begin{equation}
\hat F_{1r}\to \hat F_{1r}+\frac{3M}{r_0^2}\frac{\xi_1^r}{r_0-3M}.\label{DF1r}
\end{equation}

Putting together the results \eqref{DhR1}, \eqref{DhR2}, and \eqref{DF1r} in the formula~\eqref{Utilde_GWv2} for $\btilde U$, I determine, as desired, that $\btilde U$ is gauge-invariant:
\begin{equation}
\btilde U\to \btilde U.
\end{equation}   

Before proceeding to the next section, I note the gauge \emph{dependence} of another quantity:
\begin{equation}
\hat u^t=\hat U = U_0\left\{1+\epsilon^2\frac{r_0^2}{6M}(r_0-3M)(\hat F_{1r})^2+O(\epsilon^3)\right\}\label{U_expansion},
\end{equation}
which describes $dt/d\tau$ along $\hat z^\mu$. This expansion can be obtained from Eqs.~\eqref{U}, \eqref{r1}, and \eqref{r2}. The resulting expression is clearly gauge dependent, but only at second order. At first order, both $\hat u^t$ and $\btilde u^t$ are invariant; in the form~\eqref{U_expansion}, the first-order invariance of $\hat u^t$ is trivial (and vacuous), though it can also be seen to follow from the more meaningful fact that $\hat u^\alpha\to\hat u^\alpha-\frac{d\xi_1^\alpha}{d\tau}=\hat u^\alpha$ for a gauge vector satisfying Eq.~\eqref{xi_symmetry}. At second order, this is no longer the case, and we now see more strikingly the importance of normalizing in the effective metric, rather than the background metric, to obtain a gauge-invariant redshift.

\subsection{Transformation to an asymptotically flat gauge}\label{asymptotically_flat}

Although I have presented the formalism as if all calculations are to be performed in the Lorenz gauge, in practice that gauge must be slightly tweaked. It is known that the first-order metric perturbation $h^1_{\mu\nu}[\hat z]$ in the Lorenz gauge is not asymptotically flat, with the monopole piece of its $tt$ component approaching the constant $-2\alpha$ as $r\to\infty$; see, e.g., the discussion in Ref.~\cite{Sago-Barack-Detweiler:08}. In the notation of this paper, the constant factor is $\alpha=m/[\hat r(\hat r-3M)]^{1/2}$. To cure this ill behavior, one must perform a gauge transformation generated by 
\begin{equation}
\xi_1^\mu=-\alpha t \delta^\mu_t,\label{to-flat}
\end{equation}
which alters $h^1_{\mu\nu}$ (as well as $h^{\R1}_{\mu\nu}$) by an amount 
\begin{equation}
\Delta h^1_{\mu\nu}=2(1-2M/r)\alpha \delta_\mu^t\delta_\nu^t. 
\end{equation}
Any vector that smoothly goes to $\xi_1^\mu\to-\alpha t \delta^\mu_t$ at large $r$ would, of course, make the metric asymptotically flat, but the coefficient in front of $t$ must be constant in order to preserve the metric's helical symmetry.

Even though this transformation leaves the perturbation's helical symmetry intact, it lies outside the class of gauge vectors that satisfy Eq.~\eqref{xi_symmetry}, and $\btilde U$ is not invariant under it. To see how $\btilde U$ is altered, we can apply Eq.~\eqref{z_transformation} to the circular orbit as parametrized in Eq.~\eqref{circ_prev}. The result is $\hat z^\mu\to \hat z'^\mu$, where
\begin{equation}
\hat z'^\mu = \left\{\left[1+\e\alpha+\tfrac{1}{2}\e^2\alpha^2+O(\e^3)\right] t,\ \hat r(\e),\ \pi/2,\ \Omega t\right\}.
\end{equation}
After reparametrizing to make the worldline parameter match the new coordinate time, the orbit returns to its prior form, 
\begin{equation}
\hat z'^\mu(t',\e) = \{t',\ \hat r(\e),\ \pi/2,\ \Omega' t'\},
\end{equation}
but with an altered frequency
\begin{equation}
\Omega'=\Omega \left[1-\e\alpha+\tfrac{1}{2}\e^2\alpha^2+O(\e^3)\right].
\end{equation}
This modification of the frequency will clearly affect $\btilde U$, as can be inferred from the relation $\btilde U(\Omega)$ given in Eq.~\eqref{UvsOmega}.

Because of this, one cannot directly compare Lorenz-gauge results for $\btilde U$ to PN results. One must instead use an asymptotically flat gauge, which ensures that the time $t$ and frequency $\Omega$ have the same invariant meaning in both models: the time and frequency as measured by an inertial observer at infinity. Any asymptotically flat, helically symmetric gauge would do. All the calculations leading to Eq.~\eqref{Utilde_GWv2} in the preceding sections could have been performed in any helically symmetric gauge. The only change to the derivation is that the field equations themselves would have been modified from their form~\eqref{h1_SC}--\eqref{h2_SC}, with a different differential operator on the left-hand side and a different puncture on the right. If the gauge is sufficiently nice, those changes would not alter the forms~\eqref{h1_decomposition} and \eqref{h2_decomposition} of the metric perturbation, and the time-symmetrized effective metric could still be defined in the same way as it was in the Lorenz gauge. 

To best use the puncture already derived in the Lorenz gauge, the simplest way to construct an asymptotically flat, helically symmetric metric through second order is to use the gauge vector~\eqref{to-flat} to minimally modify the field equations and puncture from their Lorenz-gauge form. Analogously, a second-order gauge vector will be needed to make the second-order field asymptotically flat. These modifications go beyond the scope of this paper, but they will be provided in a future publication~\cite{Warburton-etal:14}

\section{Alternative definitions of conservative dynamics}\label{variants}
Thus far I have used a very particular definition of the conservative dynamics, based on the motion being geodesic in a time-symmetrized effective metric. I now consider two alternative definitions.

\subsubsection{Turning off dissipative terms in the force}
In Sec.~\ref{time-symmetrized_metric} I discussed two different procedures that would lead to a circular orbit $\hat z^\mu$: (i) constructing a time-symmetrized effective metric $\btilde g_{\mu\nu}=g_{\mu\nu}+\hat h^\R_{\mu\nu}$ and making $\hat z^\mu$ a geodesic of that metric; and (ii) using an effective metric $\btilde g_{\mu\nu}=g_{\mu\nu}+h^\R_{\mu\nu}$ constructed from the retarded, nonsymmetrized metric, and simply neglecting the dissipative, $F^t$ and $F^\phi$, components of the resulting self-force. In a given gauge and for a given orbital frequency, the orbits in methods (i) and (ii) will have different radii. 

Now suppose I had adopted option (ii) as my definition of the conservative dynamics, using the radial force~\eqref{FrV2} rather than~\eqref{hatFr}, and normalizing $\btilde u^\mu$ with respect to $\btilde g_{\mu\nu}=g_{\mu\nu}+h^\R_{\mu\nu}$ rather than $g_{\mu\nu}+\hat h^\R_{\mu\nu}$. The derivation of the formula~\eqref{Utilde_GW_prev} for $\btilde U$ in Sec.~\eqref{preview}, and the details provided in Secs.~\ref{U_SC}, \ref{expanded_motion}, and \ref{Utilde_GW} would have carried through virtually unchanged, since they did not rely on any particular definition of the radial force. The single change would have been the replacement of $\hat h^\R_{\mu\nu}$ with $h^\R_{\mu\nu}$ in the normalization of $\btilde u^\mu$. As discussed in Sec.~\ref{Utilde_GW}, the value of the formula~\eqref{Utilde_GW_prev} is identical whether  $\hat h^\R_{\mu\nu}$ or $h^\R_{\mu\nu}$ is used therein. Hence, \emph{the two definitions of the conservative dynamics yield exactly the same value of $\btilde U$}, even though the quantity $\btilde U=\btilde u^t$ in the two definitions refers to the four-velocity on slightly different circular orbits.

Only one difficulty arises in this second definition. As mentioned below Eq.~\eqref{DhR2}, if $h^\R_{\mu\nu}$ is used in the formula for $\btilde U$, then $\btilde U$, as given by Eq.~\eqref{Utilde_GW_prev}, is \emph{not} gauge invariant; its invariance is broken by the fact that $h^{\R1}_{u_0u_0,t}\neq 0\neq h^{\R1}_{u_0u_0,\phi}$. Oddly, no matter the choice of gauge, the value of $\btilde U$ is unchanged by making the replacement $\hat h^\R_{\mu\nu}\to h^\R_{\mu\nu}$---and yet the formula is invariant under gauge transformations only if the symmetrized field is used. This conundrum is resolved as follows: The formula~\eqref{Utilde_GW_prev} for $\btilde U$ utilizes an expansion of $\hat z^\mu$ in which the corrections $\hat z_1^\mu$ to $z_0^\mu$ are purely radial, but if the gauge vector $\xi^\mu_1$ has $t$ or $\phi$ components, then $\hat z^{t\ {\rm or }\ \phi}_1\to\hat z^{t\ {\rm or }\ \phi}_1 - \xi_1^{t\ {\rm or}\ \phi}$. Therefore, Eq.~\eqref{Utilde_GW_prev}, with no time symmetrization of the effective metric, is naturally valid only in a class of gauges related by gauge vectors that reduce to $\xi^\mu_n=\delta^\mu_r \xi^r_n$ on $z_0^\mu$. Within that class of transformations, $h^{\R1}_{u_0u_0,t}\xi^t_1$ and $h^{\R1}_{u_0u_0,\phi}\xi^\phi_1$ do not appear in the transformation of $\btilde U$. One could instead write a more general formula for $\btilde U$ that allows $\hat z_1^\mu$ to include arbitrary (constant) shifts $\hat t_1$ and $\hat \phi_1$ in the time and phase of the orbit. That formula would be invariant in the broader class of gauges related by transformations satisfying Eq.~\eqref{xi_symmetry}, and its numerical value would, of course, be independent of $\hat t_1$ and $\hat \phi_1$. But the necessary involvement of those arbitrary constants would be somewhat unnatural. The definition of conservative dynamics based on a geodesic in a time-symmetrized metric bypasses these issues.

\subsubsection{Standing-wave approximation}
Another way of defining conservative dynamics would be to construct a truly conservative physical system, rather than trying to extract the conservative portion of a dissipative system. This could be done by setting up standing waves, balancing the outgoing radiation with incoming radiation. Mathematically, this would correspond to adopting the half-retarded-plus-half-advanced first-order solution, using that solution within the second-order Ricci tensor in the second-order field equation~\eqref{h2_SC}, and once again adopting a half-retarded-plus-half-advanced solution. With this construction, the effective metric would automatically be time symmetric and the force purely radial. Hence, Eq.~\eqref{Utilde_GW_prev} would again apply, but the second-order regular field would differ from that used in the other definitions of conservative dynamics.

There are two reasons for not following this route: First, it would not be useful for comparing with PN theory, where conservative dynamics are always extracted from retarded solutions. Second, it would introduce additional numerical challenges, because the standing waves at infinity would lead to an infrared divergence in the second-order field. This divergence can be estimated by analyzing the behavior of the solution and the Green's function at large $r$. The first-order half-retarded-plus-half-advanced solution contains terms like $e^{ikr}/r$ and terms like $e^{-ikr}/r$ (with $k\geq0$), meaning the source $\delta^2 R_{\mu\nu}$ in the second-order field equation will contain terms like $e^{i(k_1+k_2)r}/r^2$, among others. If we write the second-order modes $h_{2i\ell m}$ as an integral over a Green's function and examine the contribution to the integral from a region of large $r$, then we can approximate the half-retarded-plus-half-advanced Green's function with that for the Helmholtz equation in flat space, $G_k(x,x')= \frac{e^{ik|\vec x-\vec x'|}+e^{-ik|\vec x-\vec x'|}}{2|\vec x-\vec x'|}$. Further specializing to $r'\gg r\gg M$, we may write $G_k(x,x')\sim \frac{e^{ikr'}}{r'}$. The contribution to the second-order solution from terms like $e^{-i(k_1+k_2)r}/r^2$ in the source will then be $\sim\int G_k(r,r')\frac{e^{i(k_1+k_2)r'}}{r'^2} r'^2dr'\sim\int \frac{e^{i(k_1+k_2-k)r'}}{r'}dr'$. This diverges as $\sim \ln r'$ when $k=k_1+k_2$.\footnote{This is also the reason why the time-symmetrized effective metric in Sec.~\ref{time-symmetrized_metric} should be considered a local construction rather than the regular piece of a half-retarded-plus-half-advanced global solution: the retarded integral of a product of advanced solutions generically diverges. So the globally symmetrized field would likely be ill behaved.} 

The ill behavior at large $r$ might be overcome, perhaps using methods devised to describe purely conservative systems in the fully nonlinear problem~\cite{Detweiler:89, Blackburn-Detweiler:92, Detweiler:93, Friedman-etal:01,Shibata-etal:04,Whelan-etal:00,Andrade-etal:03,Beetle-etal:07,Hernandez-etal:09,Gourgoulhon-etal:01,Grandclement-etal:01}. However, since the standing-wave construction is unlikely to agree with PN results, it is of limited relevance.

\section{Summary and outline of numerical scheme}\label{scheme}
The main result of this paper is Eq.~\eqref{Utilde_GW_prev}, which is an extension of Detweiler's redshift invariant $\btilde u^t\equiv\btilde U$ to second order. This formula describes the ratio between intervals of Schwarzschild coordinate time and proper time on a precisely circular orbit $\hat z^\mu$ that is accelerated only by a conservative piece of the self-force; the proper time is measured in a certain effective metric in which $\hat z^\mu$ is a geodesic. However, the formula is written in terms of quantities evaluated not on $\hat z^\mu$, but on a nearby circular orbit $z_0^\mu$, of the same orbital frequency, that is a geodesic of the background metric. 

This result utilizes the Gralla-Wald picture of perturbed motion, in which the perturbed orbit $\hat z^\mu$ is described as a deviation from a background geodesic $z_0^\mu$. Before arriving at that picture, my analysis began in a self-consistent picture, in which the orbit sourcing the metric perturbations is self-consistently accelerated by those perturbations. In that picture, I derived a formula for $\btilde U$, given by Eq.~\eqref{Utilde_SC}, in which all quantities were evaluated on the accelerated orbit. At the beginning of Sec.~\ref{GW_formalism}, I described why a self-consistent numerical scheme to calculate this quantity $\btilde U$ is not ideal: it requires one to know the orbit $\hat z^\mu$ in advance; in other words, one must determine the correct initial data for a circular orbit through second order in perturbation theory. This challenge does not arise when one works in the Gralla-Wald picture, because the background geodesic may be freely prescribed, making the Gralla-Wald picture ideal for a concrete numerical calculation of $\btilde U$. Indeed, over the course of my analysis, I have described most of the key ingredients for such a calculation. Putting those ingredients together, we arrive at the following scheme: 
\begin{enumerate}
\item Choose a circular geodesic of the background metric. This amounts to choosing an orbital radius $r_0$.
\item Assume decompositions 
\begin{align}
h^n_{\mu\nu} &= \sum_{i\ell m}h_{ni\ell m}e^{-im\Omega t}Y^{i\ell m}_{\mu\nu},\\
h^{\res n}_{\mu\nu} &= \sum_{i\ell m}h^\res_{ni\ell m}e^{-im\Omega t}Y^{i\ell m}_{\mu\nu}
\end{align}
of the retarded and residual fields, with the frequency given by Eq.~\eqref{Omega0_prev}. 
\item Solve the separated version of the first-order field equation~\eqref{h1_GW} [or~\eqref{h1_GW_point}] to obtain (i) the radial functions $h_{1i\ell m}(r)$ at all points $r\neq r_0$, and (ii) the regular field $h^{\R1}_{\mu\nu}$ and its derivatives $h^{\R1}_{\mu\nu,\rho}$ on $z_0^\mu$.  Transform these results to the asymptotically flat gauge using the gauge vector $\xi_1^\mu$, given in Eq.~\eqref{to-flat}.
\item With the (transformed) numerical values of $h^{\R1}_{\mu\nu}$ and $h^{\R1}_{\mu\nu,r}$, calculate (i) the first-order radial force, using Eq.~\eqref{F1}, (ii) the first-order conservative shift in orbital radius, $\hat r_1$, using Eq.~\eqref{r1}, and (iii) the tensor $\delta m_{\mu\nu}$, using Eq.~\eqref{dm_GW}.
\item Construct and evaluate the radial functions $S^{\rm eff}_{2i\ell m}(r)$ in the source $S^{2\rm eff}_{\mu\nu}=\sum_{i\ell m}S^{\rm eff}_{2i\ell m}e^{-im\Omega t}Y^{i\ell m}_{\mu\nu}$ for the second-order field equation. This involves
		\begin{enumerate}
			\item rewriting Eq.~\eqref{h2_GW} to account for the transformation generated by $\xi_1^\mu$,
			\item using the coupling formula~\eqref{coupling} to calculate the radial functions in the decomposition of $\delta^2 R_{\mu\nu}[h^1,h^1]$ from the radial functions $h_{1i\ell m}(r)$,
			\item constructing a puncture of the form $h^{\P2}_{\mu\nu}=\sum_{i\ell m}h^\P_{2i\ell m}e^{-im\Omega t}Y^{i\ell m}_{\mu\nu}$, which can be done by decomposing the expansion of the singular field given schematically by Eq.~\eqref{hS2_GW_schematic} and explicitly by Eq.~(144) in Ref.~\cite{Pound-Miller:14}; as input, this puncture uses the numerical values of $\hat r_1$, $\delta m_{\mu\nu}$, and $h^{\R1}_{\mu\nu}$  on $z_0^\mu$ (and potentially the derivatives of $h^{\R1}_{\mu\nu}$, depending how many orders in $|x^\alpha-z_0^\mu|$ are used in the puncture). The puncture, which was found in the Lorenz gauge, must be tweaked to account for the transformation generated by $\xi_1^\mu$.
		\end{enumerate}
\item Solve for the radial functions $h_{2i\ell m}$ and $h^\res_{2i\ell m}$ in the separated version of the second-order field equation. \item Find a gauge vector $\xi_2^\mu$ that brings $h^2_{\mu\nu}$ to an asymptotically flat (still helically symmetric) form, and apply the resulting transformation to $h^{\R2}_{\mu\nu}$. The only necessary output from the result is $h^{\R2}_{u_0u_0}$. 
\item Combine  $h^{\R1}_{u_0u_0}$, $F^r_1$, and $h^{\R2}_{u_0u_0}$ in Eq.~\eqref{Utilde_GWv2} to calculate the redshift variable $\btilde U$.
\end{enumerate}
The technical details of this scheme, particularly those involved in steps 5 and 6, will be presented in a future paper~\cite{Warburton-etal:14}. 

A comparison of the numerically calculated $\btilde U$ to its value in PN theory will be the first test of the second-order self-force formalism. Assuming that test is passed, second-order results can begin to inform high-order PN theory and EOB. And although I have focused on a means of calculating $\btilde U$ and nothing else, the general formalism I have presented, and the same type of numerical scheme, can be used to calculate any other conservative effects that may occur on circular orbits. Most significantly, it should be straightforward to generalize the techniques of Ref.~\cite{Isoyama-etal:14} to derive a formula for the second-order shift in the frequency of the ISCO.

\begin{acknowledgments}
The formulation of this paper owes much to several of my fellow researchers. Alex Le Tiec steered me toward analyzing $\btilde u^t$ instead of $h^\R_{uu}$, and both he and Luc Blanchet provided illuminating explanations of how $\btilde u^t$ is defined in post-Newtonian theory. Leor Barack offered helpful comments on a draft of this paper, derived the relationship between modes of retarded and advanced metric perturbations, and aided me enormously in understanding their asymptotic behaviors. I also thank Niels Warburton for numerically confirming properties of first-order Lorenz-gauge solutions, and Jonathan Thornburg, Soichiro Isoyama, Sam Dolan, Sarp Akcay, and Eric Poisson for helpful discussions. The research leading to these results received funding from the European Research Council under the European Union's Seventh Framework Programme (FP7/2007-2013)/ERC Grant No. 304978.
\end{acknowledgments}

\appendix

\section{Gralla-Wald picture including dissipation}\label{dissipation}
In Sec.~\ref{Utilde_GW}, I derived the gauge-invariant quantity $\btilde U$ in the Gralla-Wald picture by starting with the self-consistent equation of motion for the circular orbit $\hat z^\mu$. In this section I show how the Gralla-Wald picture looks when dissipation is accounted for. Rather than expanding $\hat z^\mu$, I expand the physical, inspiraling orbit $z^\mu$, and from the result I construct $\hat z^\mu$ as a certain piece of the perturbative expansion. An expansion of $\btilde U$ follows naturally.

I begin by rewriting the equation of motion~\eqref{SC_motion_prev} in terms of derivatives with respect to $t$ rather than $\tau$. The result is
\begin{equation}
\frac{d^2 z^\mu}{dt^2}+U^{-1}\frac{dU}{dt}\frac{dz^\mu}{dt}+\Gamma^\mu_{\alpha\beta}(z(t))\frac{dz^\alpha}{dt}\frac{dz^\beta}{dt} = U^{-2}F^\mu,\label{self-consistent_t}
\end{equation}
where
\begin{equation}
U\equiv \frac{dt}{d\tau}.
\end{equation}

I next expand $z^\mu(t,\e)$ as
\begin{equation}
z^\mu(t,\e) = z_0^\mu(t) +\e z^\mu_1(t)+\e^2 z^\mu_2(t)+O(\e^3),\label{expanded_z}
\end{equation}
where $z_0^\mu(t)=\{t,r_0,\pi/2,\Omega_0 t\}$ and the perturbations are given by
\begin{equation}
z_1^\mu(t)=\{0,r_1(t),0,\phi_1(t)\}
\end{equation}
and the analogue for $z_2^\mu$. By substituting this expansion into the normalization condition $g_{\mu\nu}(z)u^\mu u^\nu=-1$, one obtains
\begin{align}
U^{-2} &= U_0^{-2} - 2\epsilon \Omega_0 r_0^2 \dot\phi_1 + \epsilon^2\bigg\{-\frac{3 M r_1^2}{r_0^3} - 4 \Omega_0 r_0 r_1 \dot\phi_1 \nonumber\\
&\quad
 - f_0^{-1}(\dot r_1)^2 - r_0^2 \left[(\dot\phi_1)^2+2\Omega_0 \dot\phi_2\right]\bigg\} + O(\epsilon^3),\label{U_with_diss}
\end{align}
where an overdot denotes a derivative with respect to $t$.

Now inserting \eqref{expanded_z} into Eq.~\eqref{self-consistent_t} yields 
\begin{align}
\frac{2M}{r_0^2f_0}\dot r_1+\frac{M}{\Omega_0(r_0-3M)}\ddot\phi_1 &= F^t_1 (1-3M/r_0),\label{phi1_equation}\\
\ddot r_1 - \frac{3M}{r_0^3}f_0 r_1 - 2r_0f_0\Omega_0 \dot\phi_1 &= F_1^r(1-3M/r_0).\label{r1_equation}
\end{align}
$F^\mu_1$ is given by the first-order terms in Eqs.~\eqref{FtV1} and \eqref{FrV1}, with $h^\R_{\mu\nu}[\hat z]$ replaced by $h^{\R1}_{\mu\nu}[z_0]$; note that this force is constant along $z_0^\mu$. Although a symbolic mathematics package such as Mathematica will readily provide the general solution to the system~\eqref{phi1_equation}--\eqref{r1_equation}, solving it manually will be instructive. I first integrate Eq.~\eqref{phi1_equation} once to find
\begin{align}
\dot\phi_1(t) &= \dot\phi_1(0) + tF^t_1 \frac{r_0}{M}\Omega_0 (1-3M/r_0)^2 \nonumber\\
&\quad- \frac{2\Omega_0}{r_0f_0}(1-3M/r_0)[r_1(t)-r_1(0)].\label{phi1dot}
\end{align}
Equation~\eqref{r1_equation} then becomes a formula for a forced harmonic oscillator, 
\begin{equation}
\ddot r_1 +\Omega_0^2 (1-6M/r_0)r_1 = A+Bt,\label{r1_equationV2}
\end{equation}
where the driving terms are
\begin{align}
A &\equiv \left[F^r_1+\frac{4M}{r_0^3}r_1(0)\right](1-3M/r_0)\nonumber\\
	&\quad +2r_0 f_0 \Omega_0\dot\phi_1(0)\\
B &\equiv \frac{2}{r_0}f_0(1-3M/r_0)^2 F_1^t. 
\end{align}
The general solution to Eq.~\eqref{r1_equationV2} can be found by the method of variation of parameters, which yields 
\begin{align}
r_1(t) &= C_1\cos\omega t +C_2\sin\omega t + \frac{A}{\omega^2}(1-\cos\omega t)\nonumber\\
&\quad +\frac{B}{\omega^2}\left(t-\frac{1}{\omega}\sin\omega t\right),
\end{align}
with an oscillation frequency given by
\begin{equation}
\omega\equiv \Omega_0\sqrt{1-6M/r_0}.
\end{equation}

The oscillatory terms that do not depend on the self-force correspond to a small shift away from a circular geodesic toward an eccentric one, in which the radius oscillates with time. Notice that for $r_0<6M$, the frequency of the oscillations became imaginary, and oscillations became exponential growth. This corresponds to the fact that $r_0=6M$ is the innermost stable circular orbit; at smaller radii, the zeroth-order geodesic is unstable, and there is no ``nearby" eccentric geodesic to perturb toward.

I wish to describe a situation in which the orbit is circular if the self-force vanishes. Hence, I wish to remove the perturbations toward an eccentric geodesic. This is accomplished by choosing $C_1=A/\omega^2$ and $C_2=B/\omega^3$, leading to
\begin{equation}
r_1(t) = \frac{A}{\omega^2}+\frac{B}{\omega^2}t.
\end{equation}
This choice constrains the choice of initial conditions. From $r_1(0)=A/\omega^2$, I find
\begin{equation}
r_1(0) = -\frac{1-6M/r_0}{3f_0\omega^2}\left[F^r_1(1-3M/r_0)+2r_0 f_0 \Omega_0\dot\phi_1(0)\right]
\end{equation}
We are still left with the freedom to choose either $r_1(0)$ or $\dot\phi_1(0)$. I choose $\dot\phi_1(0)=0$, such that
\begin{equation}
r_1(0) = -\frac{r_0^2(r_0-3M)}{3Mf_0}F_1^r.
\end{equation}
This is equal to the first-order term $\hat r_1$ in the expansion of $\hat z^\mu$, as we saw previously in Eq.~\eqref{r1}.

Equation~\eqref{phi1dot} can now be straightforwardly integrated to find $\phi_1(t)$. My final results for the first-order corrections to the inspiraling worldline are
\begin{align}
r_1(t) &= -\frac{r_0^2(r_0-3M)}{3Mf_0}F_1^r + \frac{2r_0f_0(r_0-3M)^2}{M(r_0-6M)}F_1^t t \label{r1_diss}\\
\phi_1(t) &= -\frac{3f_0(r_0-3M)^2}{2M(r_0-6M)} \Omega_0 F_1^t t^2. \label{phi1_diss}
\end{align}

One can clearly see from these results why the Gralla-Wald picture is ill suited to treating dissipation: the corrections $r_1$ and $\phi_1$, which are assumed to be small, grow large with time. At higher orders, the corrections will grow even more rapidly. But we can nevertheless extract the conservative dynamics. The first-order term in the conservatively accelerated worldline $\hat z^\mu=z_0^\mu(t)+\e \hat z_1^\mu+\ldots$ is found simply by turning off $F^t_1$ and $F^\phi_1$, leaving only the constant correction $r_1(t)=r_1(0)=\hat r_1$. Figure~\ref{orbits} displays the three orbits $z^\mu=z_0^\mu+\e z_1^\mu$, $\hat z^\mu=z_0^\mu(t)+\e \hat z_1^\mu$, and $z_0^\mu$.

This same procedure could be carried to second order, but including all dissipative effects would require some knowledge of the time dependence of the second-order force. We can, however, find the second-order term in $\hat z^\mu$ by solving the second-order term in Eq.~\eqref{self-consistent_t} with $F_n^t$ and $F_n^\phi$ set to zero everywhere. Doing so recovers Eq.~\eqref{r2} for the second-order conservative correction to the radius and zero for all other second-order terms in $\hat z^\mu$. (I gloss over the question of whether the radial force is that corresponding to a time-symmetrized effective metric or not, as per the discussion in Sec.~\ref{variants}.) Equation~\eqref{U_expansion} for $\btilde U$ can then be found by calculating $\frac{dt}{d\btilde\tau}$ directly from the perturbative expansion $\hat z^\mu(t,\e)=z_0^\mu(t)+\delta^\mu_r (\e \hat r_1+\e^2 \hat r_2)$.

\bibliography{../../bibfile}
\end{document}